\documentclass[11pt]{article}
\usepackage[preprint]{acl}

\usepackage{times}
\usepackage{latexsym}
\usepackage[T1]{fontenc}
\usepackage[utf8]{inputenc}
\usepackage{microtype}
\usepackage{inconsolata}
\usepackage{graphicx}
\usepackage{amsmath,amssymb}
\usepackage{booktabs,longtable,array,calc}
\usepackage{tabularx}
\usepackage{url}
\usepackage{xcolor}
\usepackage{iftex}
\usepackage{placeins}
\usepackage{multicol}
\usepackage{dblfloatfix}
\ifXeTeX
  \usepackage{xeCJK}
\fi
\providecommand{\real}[1]{#1}

\newcommand{\numcite}[1]{\textsuperscript{#1}}
\title{Unfolding the Interdisciplinary Complexities of Climate Science:\\
Fuxi-Climate Foundational Model}
\author{
\textbf{Zhengyu Shi\textsuperscript{a,b,i,j,1}, Shaojie Shi\textsuperscript{b,i,1},
Rui Xu\textsuperscript{b,i}, Bohao Lv\textsuperscript{b,i}}\\
\textbf{Zhichao Chen\textsuperscript{c}, Jiaran Hao\textsuperscript{b},
Zijian Chen\textsuperscript{d}, Weiqi Tang\textsuperscript{e,f}}\\
\textbf{Yuan Qi\textsuperscript{a,b,i,j}, Yinghui Xu\textsuperscript{b,*},
Libo Wu\textsuperscript{a,g,h,i,j,*}}\\
\textsuperscript{a}College of Computer Science and Artificial Intelligence, Fudan University\\
\textsuperscript{b}Artificial Intelligence Innovation and Incubation (Al$^3$) Institute, Fudan University\\
\textsuperscript{c}Department of Environmental Science and Engineering, Fudan University\\
\textsuperscript{d}School of Data Science, Fudan University\\
\textsuperscript{e}Fudan Development Institute, Fudan University\\
\textsuperscript{f}Shanghai Institute for Energy and Carbon Neutrality Strategy\\
\textsuperscript{g}Institute for Big Data, Fudan University\\
\textsuperscript{h}MOE Laboratory for National Development and Intelligent Governance\\
\textsuperscript{i}Shanghai Innovation Institute\\
\textsuperscript{j}Shanghai Academy of AI for Science\\
\textsuperscript{1}Equal contribution; \textsuperscript{*}Correspondence:
\texttt{wulibo@fudan.edu.cn}, \texttt{xuyinghui@fudan.edu.cn}
}

\begin{document}
\maketitle
\begin{abstract}
Climate research and decision-making require integrating evidence across physical processes, socio-economic dynamics and policy responses. Large language models (LLMs) have been explored for accessing and synthesizing climate knowledge, but their ability to support structured interdisciplinary reasoning is still limited. Here we present the Fuxi-Climate Foundation Model (CFM), a climate-specialized LLM designed to support consistent reasoning across domains. CFM maintains more stable analytical behavior as interdisciplinary complexity increases, whereas performance in other models becomes more variable. On expert-designed climate transition tasks, CFM produces more structured analyses that explicitly address trade-offs and uncertainty, achieving 45\% trade-off coverage and 47.27\% uncertainty-aware reasoning. These results indicate that CFM can support more realistic analysis of climate risks and transition pathways, and provide a basis for agent-based systems to explore complex policy and decision scenarios. The model is openly available at \url{https://huggingface.co/SII-yuning/cfm}.
\end{abstract}

\addtocontents{toc}{\protect\setcounter{tocdepth}{-1}}
\section{Introduction}
Addressing climate change requires combining insights from multiple domains, including physical climate processes\numcite{1,2}, socio-economic dynamics\numcite{3,4} and policy decision-making\numcite{5,6}. Such questions, including the evaluation of climate risks and the design of mitigation and adaptation strategies, cannot be resolved within a single disciplinary framework, but require the integration of interdisciplinary evidence and reasoning\numcite{7,8}. Collective assessment frameworks have long played a central role in addressing this challenge. The Intergovernmental Panel on Climate Change (IPCC), for example, synthesizes evidence across disciplines by coordinating large expert communities and producing periodic assessment reports\numcite{9}. This process provides a foundation for scientific consensus and supports global climate governance\numcite{10,11}. However, this assessment is shaped by its periodic and consensus-based structure, which may limit the timely incorporation of emerging findings and rapidly evolving practices\numcite{12,13}.

Recent advances in large language models (LLMs) have led to growing interest in their use for climate science and decision-making. Previous studies have explored their application in settings such as grounded question-answering linked to authoritative climate sources\numcite{14}, synthesis of adaptation\numcite{15} and mitigation\numcite{16} evidence, and analysis of policy and governance texts\numcite{17,18}. Across these efforts, LLMs have shown the ability to reduce the effort required to navigate fragmented bodies of scientific and institutional knowledge and to support exploratory analysis across domains. Similar approaches have also been proposed for public-facing climate services\numcite{19,20}, extending access to climate-relevant information beyond expert communities.

However, these capabilities do not readily translate into the ability to conduct climate analysis. Although LLMs can generate fluent and informative responses\numcite{21}, empirical evaluations show that their performance becomes inconsistent when tasks require integration across disciplines and sustained reasoning beyond surface-level summaries\numcite{22}. This limitation becomes more evident when reasoning unfolds across multiple analytical steps (Fig. 1). While many models can identify problem structures in initial stages, their performance declines during interdisciplinary analysis, and only a small fraction of responses proceed to coherent climate transition pathway design or policy formulation, with fewer than 20\% reaching this stage.

\begin{figure*}[!t]
  \centering
  \includegraphics[width=.88\textwidth,height=.50\textheight,keepaspectratio]{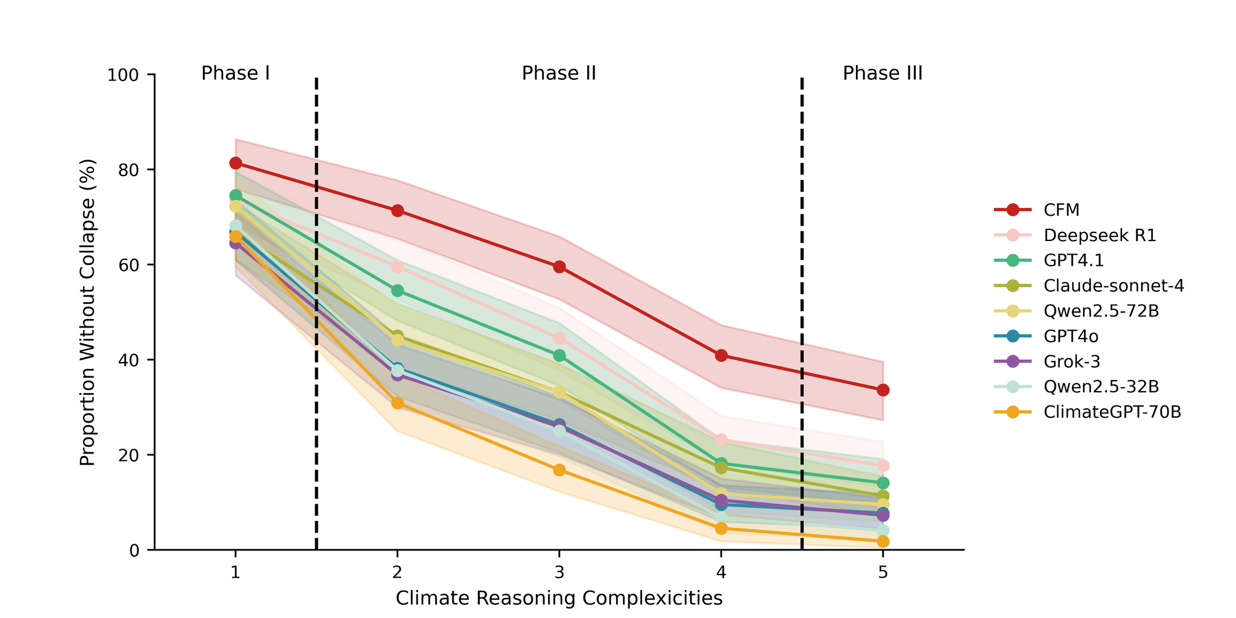}
  \caption{\textbf{Stability of climate reasoning with increasing analytical depth.} The x-axis represents successive stages of climate transition reasoning, from problem identification (Phase I) to interdisciplinary analysis (Phase II) and transition design (Phase III), reflecting increasing depth of reasoning. The y-axis shows the proportion of responses that remain logically consistent without breakdown during multi-step reasoning. As reasoning deepens, general-purpose language models exhibit a rapid decline in stability, whereas the Fuxi-Climate Foundation Model (CFM) maintains higher consistency, particularly in later stages requiring interdisciplinary integration.}
  \label{fig:1}
\end{figure*}

These patterns likely reflect limitations in how current models organize and deploy knowledge. Because training is primarily driven by next-token prediction, models are optimized for generating text that is statistically plausible, rather than for maintaining causal consistency or structured reasoning across domains\numcite{23,24}. As a result, they may struggle to preserve relationships across steps or to respect implicit constraints when reasoning about complex climate systems\numcite{25}. This challenge becomes more apparent in tasks that require integrating physical processes with socio-economic dynamics and policy considerations, or balancing multiple objectives under uncertainty\numcite{26}. These observations suggest that current approaches improve access to climate knowledge without fully addressing the coordination of reasoning across disciplines.

To address this gap, we develop the Fuxi-Climate Foundation Model (CFM), a climate-specialized large language model designed to support structured reasoning across disciplines. Our approach supports structured reasoning across physical, socio-economic and policy domains. This is achieved through a unified framework that enables interdisciplinary climate reasoning to be systematically learned, aligned and evaluated. The model is exposed to diverse climate knowledge through large-scale cross-domain training, further shaped by expert-designed instructions that capture how climate scientists structure problems and develop analyses, and assessed using a benchmark that explicitly tests the consistency of reasoning under increasing interdisciplinary complexity.

We evaluate the model across multiple levels of complexity. On structured climate benchmarks, performance diverges across models as interdisciplinary complexity increases, while the climate foundation model maintains more consistent analytical behavior across disciplines. In cross-disciplinary settings, the model shows improved robustness as the number of domains involved in a task increases. On expert-designed, open-ended climate transition problems, the model produces more structured analyses that explicitly address trade-offs and uncertainty. It achieves 45\% coverage of explicit trade-off analysis and 47.27\% coverage of uncertainty-aware reasoning, substantially exceeding existing models. These improvements are observed across core climate domains, including mitigation, adaptation and climate--health--food systems. These capabilities indicate that the model can support more realistic climate analysis, where decisions must balance competing objectives under uncertainty. They also enable CFM to serve as a foundation for agent-based climate systems, enabling coherent multi-step reasoning and interaction in complex analytical workflows. These results suggest that CFM can serve as a new research infrastructure for climate science, helping researchers synthesize fragmented knowledge, interrogate transition pathways and explore complex policy questions in a more integrated way.

\section{Results}

\subsection{Interdisciplinary Knowledge Foundation for Climate Science}

We construct a unified interdisciplinary knowledge foundation for climate science (Fig. 2a). This foundation integrates three components corresponding to different stages of model development: a climate corpus for continued pre-training (CPT), ClimaInstruct for supervised fine-tuning (SFT), and ClimaBench for evaluating climate understanding and reasoning consistency.

We first develop a climate corpus comprising approximately 78 billion tokens to represent knowledge across natural and socio-economic systems. The corpus integrates authoritative sources, including assessment reports from international organizations such as the IPCC, together with academic literature, policy documents, technical reports and climate-relevant web texts (see Supplements~B1--B3). This collection provides broad coverage of climate-related concepts spanning physical sciences, energy systems, economic processes and governance contexts, forming a structured basis for domain-specific knowledge acquisition during pre-training.

We then develop ClimaInstruct to align models with the interdisciplinary reasoning structure of climate science. The dataset contains 19,110 expert-validated instructions, distilled from an initial pool of 72,000 candidates through multi-stage filtering (26.54\% retained). It spans 19 climate-relevant disciplines and 144 subfields based on a standardized academic classification scheme (Fig. 2b, see Supplements~A, C3, and C4). Rather than focusing on single-domain queries, ClimaInstruct emphasizes cross-disciplinary integration, with many instructions involving multiple fields (Fig. 2c). For example, 93.46\% of instructions are associated with at least two disciplines, and 18.01\% involve five or more. This design reflects the interconnected nature of climate problems, where meaningful analysis often requires linking physical processes with socio-economic impacts and policy contexts.

To evaluate these capabilities, we further construct ClimaBench, a benchmark comprising 7,777 expert-designed tasks that span diverse forms of climate inquiry. The benchmark covers multiple task formats and levels of domain specificity (see Supplements~D1 and D3), and is annotated using the same classification scheme as ClimaInstruct (Fig. 2d). A substantial proportion of questions (74.42\%) involve two or more disciplines (Fig. 2e). ClimaBench provides a standardized evaluation of climate knowledge and reasoning across natural and socio-economic domains, enabling systematic assessment of models' ability to integrate knowledge and perform interdisciplinary reasoning in complex climate scenarios.

These components establish an interdisciplinary knowledge foundation that supports both the training and evaluation of climate-focused language models, and provide a structured basis for examining their ability to perform integrated socio-climate reasoning.

\begin{figure*}[!t]
  \centering
  \includegraphics[width=.88\textwidth,height=.50\textheight,keepaspectratio]{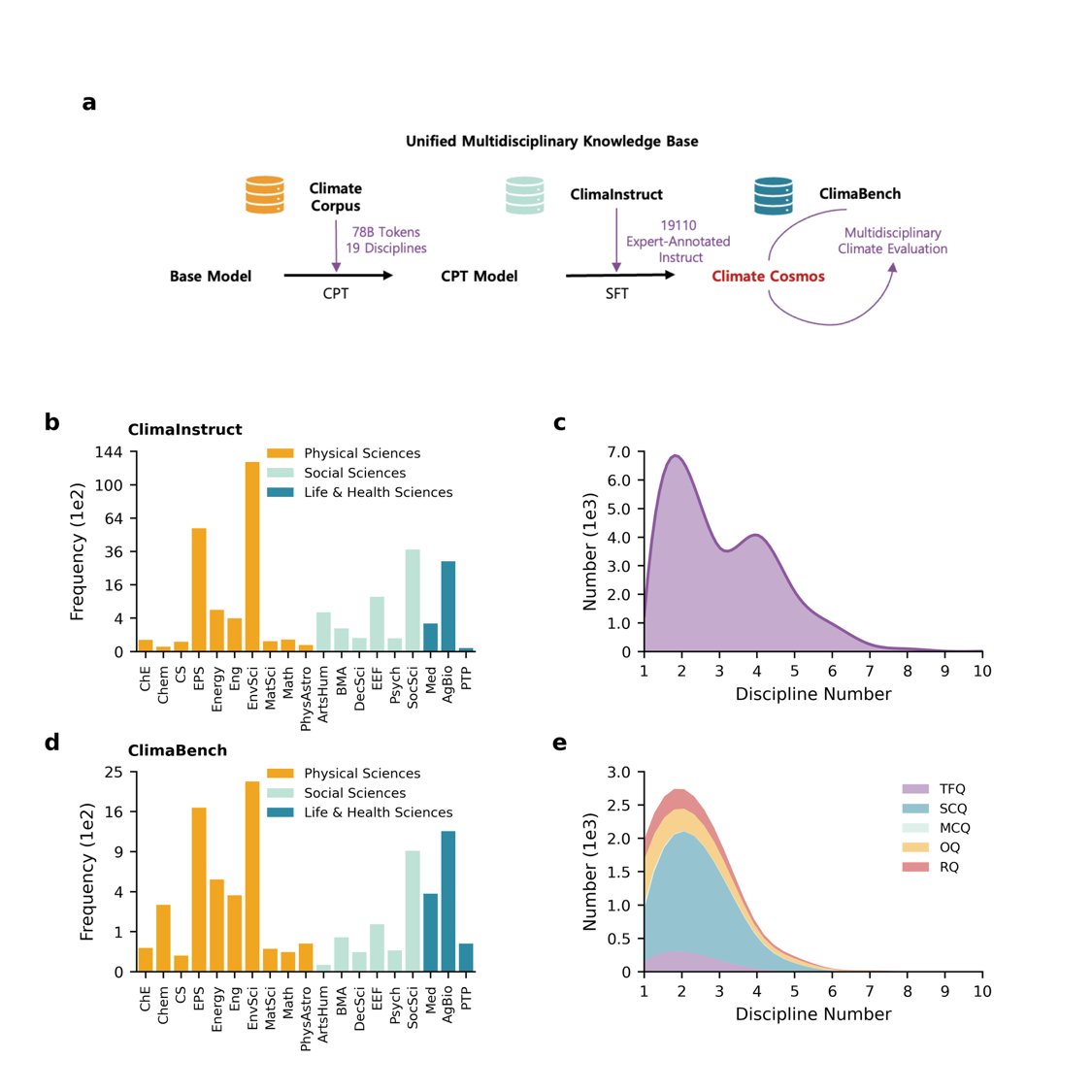}
  \caption{\textbf{Interdisciplinary Knowledge Foundation for Climate Science. a,} Schematic overview of the unified knowledge base, illustrating its composition and role in model training and evaluation. \textbf{b,} Disciplinary coverage of ClimaInstruct across climate-related fields. \textbf{c,} Interdisciplinarity of ClimaInstruct, showing the number of disciplines associated with each instruction. \textbf{d,} Disciplinary coverage of ClimaBench tasks. \textbf{e,} Interdisciplinarity of ClimaBench tasks, showing the distribution of the number of disciplines involved in each benchmark tasks.}
  \label{fig:2}
\end{figure*}

\subsection{Diverging Performance under Interdisciplinary Complexity}

We evaluate model performance on ClimaBench to examine how large language models handle climate tasks with increasing disciplinary dimensionality and task complexity (Fig. 3). Across all 19 disciplines spanning physical sciences, social sciences and life and health sciences, the climate foundation model (CFM) shows consistent performance gains relative to both general-purpose and domain-specific baselines (Fig. 3a). Performances are particularly pronounced in key climate-relevant disciplines, including Earth and Planetary Sciences (85.74\%), Environmental Science (86.27\%), Materials Science (90.14\%), Physics and Astronomy (87.34\%), and Agricultural and Biological Sciences (85.30\%), suggesting stronger capability in integrating knowledge across coupled climate--environment--technology systems.

When aggregated by levels of domain specificity, this advantage becomes more evident (Fig. 3b). On 5,450 general-level tasks, CFM achieves an average accuracy of 82.79\%, while on 2,327 specialized tasks requiring deeper domain knowledge and structured reasoning, it attains 78.06\%. In both settings, CFM outperforms leading general-purpose models, including GPT-4o, Qwen2.5-32B and LLaMA-3.1-70B, and exceeds the domain-specific baseline ClimateGPT-70B by 13.99\% and 20.20\%, respectively. Notably, the performance gap widens for specialized tasks, indicating that the model's advantage is amplified under higher levels of domain specificity and reasoning complexity (see Supplement~G1).

To further examine the effect of interdisciplinary dimensionality, we conduct a stress test by grouping tasks according to disciplinary breadth (Fig. 3c). As the number of disciplines involved in each task increases, model performance exhibits distinct scaling behaviors, reflecting differences in how models handle interdisciplinary knowledge integration. Compared with ClimateGPT-70B, CFM achieves up to a 19.71\% improvement on general-level tasks (breadth = 5), and up to a 29.00\% improvement on specialized tasks (breadth = 2), with gains remaining substantial (25.00\%) even at the highest levels of disciplinary breadth (see Supplement~G2).

Importantly, CFM exhibits a markedly stronger capacity to sustain and even improve performance as interdisciplinary complexity increases. On specialized tasks, its performance increases by 19.12\% with growing disciplinary breadth, whereas the base model shows only a marginal gain of 4.84\%. This divergence indicates that the model does not merely retain knowledge across disciplines, but is able to leverage interdisciplinary interactions to enhance reasoning outcomes. The observed scaling behavior under increasing dimensionality suggests that structured climate-specific training enables synergistic reasoning across domains, providing robustness in complex socio-climate analysis. Additional ablation and representation analyses further support this interpretation, suggesting improved cross-disciplinary alignment in the foundation model (see Supplements~E4 and F1--F3).

\begin{figure*}[!t]
  \centering
  \includegraphics[width=.88\textwidth,height=.50\textheight,keepaspectratio]{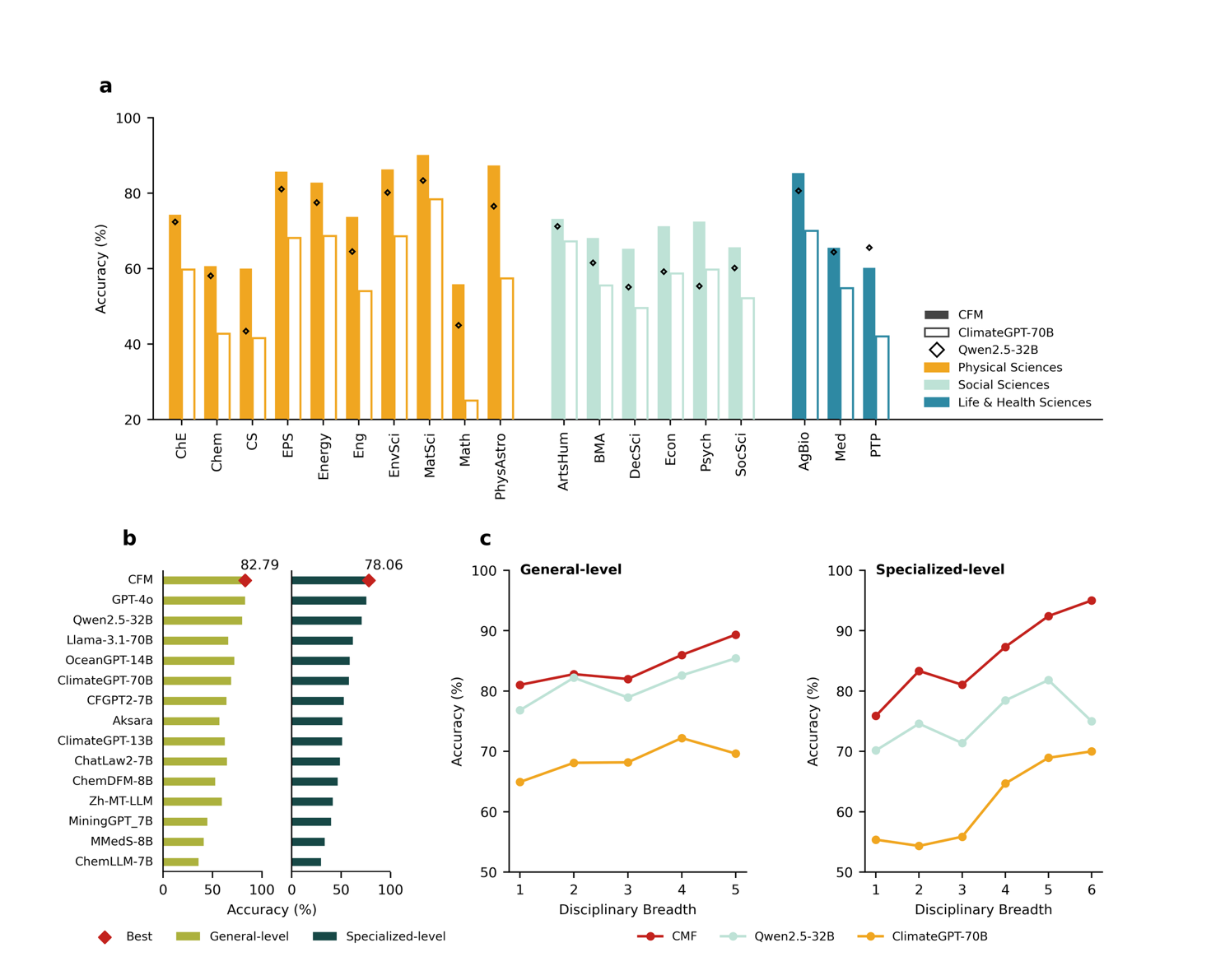}
  \caption{\textbf{Interdisciplinary Performance on Climate Tasks. a,} Disciplinary performance on ClimaBench. ClimateGPT-70B, a representative open-sourced climate model, is chosen as the primary domain-specific comparison model. Diamond markers indicate the base model (Qwen2.5-32B). \textbf{b,} Models performance on general and specialized tasks. Models include general-purpose language models and climate-related domain models (see Supplement~G3). Models are ranked by specialized-level accuracy. \textbf{c,} Interdisciplinary stress test across general and specialized tasks. Performance is evaluated as a function of disciplinary breadth, defined as the number of disciplines involved in each task. Higher accuracy indicates stronger ability to integrate knowledge across disciplines, while performance degradation with increasing breadth reflects reduced robustness under interdisciplinary complexity.}
  \label{fig:3}
\end{figure*}

\subsection{Climate Transition Pathway Analysis Aligned with Scientific Consensus}

To evaluate whether models can support realistic climate transition analysis beyond fixed-answer benchmarks, we construct 220 expert-designed tasks spanning key domains of climate mitigation, adaptation and socio-economic transformation (Fig. 4). While ClimaBench evaluates structured climate reasoning, these tasks probe model behavior in more open-ended and decision-oriented settings. These tasks do not admit a single correct answer, but instead require constructing coherent analyses that integrate evidence across disciplines and consider competing objectives under uncertainty.

We first evaluate whether model-generated analyses align with established scientific frameworks. We identify the most relevant IPCC text segments for each task and compare them with model responses. Across Top-3 to Top-20 retrieval settings, the CFM consistently achieves the highest semantic similarity (Fig. 4b), with a peak score of 0.3526, significantly exceeding the domain baseline ClimateGPT-70B (t-test, p = 0.08). Although similarity decreases as the retrieval range expands, the relative advantage of CFM remains stable, suggesting more consistent alignment with scientific consensus across retrieval settings. This pattern is further supported by improved coverage of IPCC glossary terms (see Supplement~H5), indicating broader alignment with core climate concepts.

Beyond alignment, effective climate transition analysis requires integrating knowledge across multiple disciplines. CFM generates responses that span a broader set of disciplines, with a median disciplinary breadth of 9, which is two to three disciplines higher than that of comparison models, and a pronounced long-tail distribution (Fig. 4a). This long-tail pattern indicates that a substantial proportion of responses simultaneously engage multiple climate subfields, reflecting an increased capacity to address complex problems that require coordinated reasoning across domains. In contrast, other models produce more concentrated distributions, suggesting narrower and more repetitive disciplinary perspectives.

Consistent with this pattern, CFM produces more complete analytical structures in climate reasoning. CFM achieves a reasoning step coverage of 77.27\% (Fig. 4c), indicating more systematic inclusion of key components such as problem framing, mechanism analysis and pathway design. This advantage persists across all seven climate domains (Fig. 4d). The largest gains are observed in Climate--Health--Food Systems and Mitigation \& Energy Transitions, with coverage increases of 35.71\% and 29.47\%, respectively, compared with ClimateGPT-70B.

A defining feature of climate decision-making is the need to balance competing objectives. CFM demonstrates stronger capability in trade-off analysis, explicitly comparing alternative pathways and evaluating their consequences. This occurs in 45\% (99 out of 220) of tasks, outperforming state-of-the-art models including GPT-4.1 and DeepSeek-R1 (Fig. 4e). This advantage is also observed in specific domains (Fig. 4f), with improvements of up to 61.90\% over the base model in Climate System Dynamics and Earth Processes, and up to 106.67\% over ClimateGPT-70B in Adaptation \& Climate Risk Management.

Climate transition pathways are inherently uncertain, requiring careful identification of risks and robust strategies. CFM more frequently recognizes sources of uncertainty and incorporates them into its reasoning, achieving this in 47.27\% (104 out of 220) of tasks and ranking highest among all models (Fig. 4g). This capability is consistently observed across domains (Fig. 4h), with improvements reaching up to 1.5 times that of the base model in Equity, Justice \& Societal Transformation, and up to 4 times that of ClimateGPT-70B in Mitigation \& Energy Transitions.

These results demonstrate that the climate foundation model extends beyond factual knowledge and task-level performance, enabling structured, multi-objective and uncertainty-aware analysis that more closely resembles expert reasoning in climate transition studies.

\begin{figure*}[!t]
  \centering
  \includegraphics[width=.88\textwidth,height=.50\textheight,keepaspectratio]{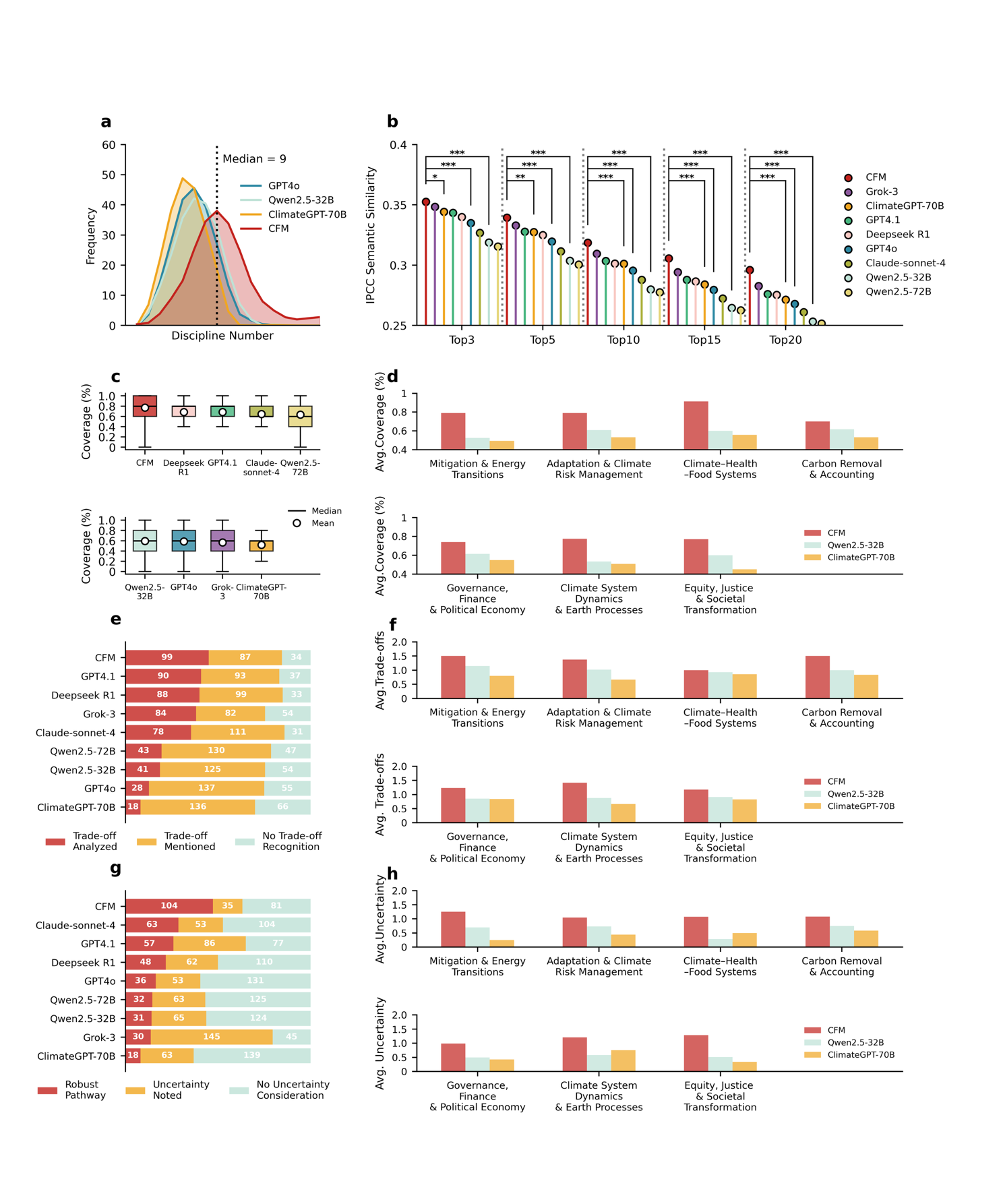}
  \caption{\textbf{Evaluating climate transition analysis in open-ended tasks.} Performance on 220 expert-designed open-ended climate transition tasks. \textbf{a,} Distribution of disciplinary breadth in model responses, measured as the number of disciplines involved in each answer. \textbf{b,} Semantic similarity between model responses and IPCC reports, computed using top-K matching text segments. Statistical significance is assessed using two-sided t-tests (*p \textless{} 0.1, **p \textless{} 0.05, ***p \textless{} 0.01). \textbf{c,} Coverage of reasoning steps in climate transition analysis, evaluated as the proportion of key analytical components present in each response. \textbf{d,} Topic-wise coverage of reasoning steps across major climate transition domains. \textbf{e,} Trade-off analysis scores, measuring the extent to which models identify and compare competing objectives in transition pathways. \textbf{f,} Topic-wise trade-off scores across climate transition domains. \textbf{g,} Uncertainty analysis scores, evaluating how models acknowledge and incorporate uncertainty in climate transition. \textbf{h,} Topic-wise uncertainty scores across climate transition domains.}
  \label{fig:4}
\end{figure*}

\section{Discussion}

Our results show that, in complex cross-disciplinary climate problems, the climate foundation model supports more consistent and structured reasoning than general-purpose language models. As interdisciplinary complexity increases, performance diverges across models, while the CFM remains more stable. This difference is particularly relevant in climate science, where tasks such as pathway design, risk assessment and policy evaluation require integrating physical processes, socio-economic dynamics and policy considerations within a unified analytical framework. This is reflected in its alignment with IPCC-style analytical structures and its improved handling of trade-offs and uncertainty, both of which are central to expert climate analysis.

From an artificial intelligence perspective, these findings suggest that the key limitation of current large language models lies not in knowledge coverage, but in their ability to support structured reasoning in open-ended, real-world problem settings. Evidence from the expert-designed tasks shows that general-purpose models struggle to construct coherent analyses involving trade-offs and uncertainty, whereas the CFM more consistently produces structured, multi-objective and uncertainty-aware reasoning. This indicates that performance in complex scientific domains depends not only on access to knowledge, but on the ability to organize and deploy that knowledge within a coherent analytical process, highlighting the importance of domain-aligned training that reflects the structure of scientific reasoning rather than relying solely on scaling.

\begin{figure*}[!t]
  \centering
  \includegraphics[width=.88\textwidth,height=.50\textheight,keepaspectratio]{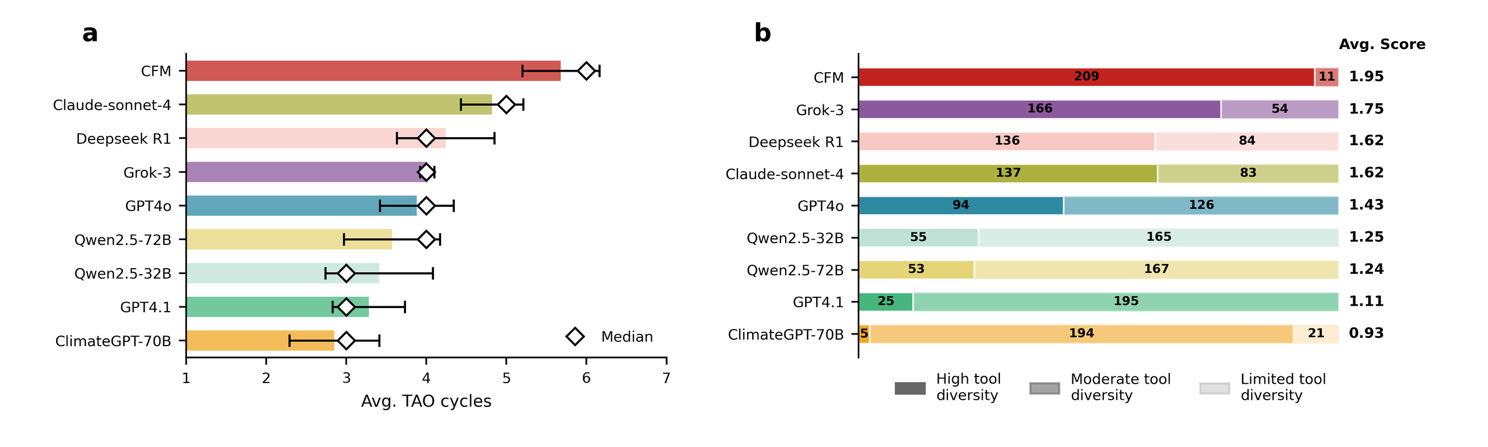}
  \caption{\textbf{Multi-step analytical behavior in agent-based climate reasoning workflows.} Evaluation on 220 expert-designed climate transition tasks within a multi-step agent reasoning framework. The agent-based workflow and evaluation protocol are described in the Supplementary Information, Supplements~I1--I3. \textbf{a,} Number of Thought--Action--Observation (TAO) reasoning loops per task, reflecting the extent to which models sustain multi-step analytical processes. \textbf{b,} Diversity of analytical actions invoked during reasoning, measured by the range and composition of tool-use behaviors, including information retrieval, mechanism analysis, scenario comparison, quantitative calculation and impact evaluation.}
  \label{fig:5}
\end{figure*}

This capability further suggests that the CFM provides a suitable basis for agent-based climate analysis. As shown in Fig. 5, the model sustains more coherent multi-step reasoning processes and exhibits greater diversity of analytical actions in open-ended tasks. This enables it to maintain structured analytical workflows required for iterative problem solving, making it well suited to serve as a reasoning backbone within agent-based systems for complex climate applications.

In real-world climate transition settings, decision-making often involves heterogeneous actors, competing objectives and evolving system dynamics that cannot be fully captured within a single analytical process. Our results show that the CFM provides a necessary basis for structured and interdisciplinary reasoning in such contexts. Building on this foundation, future work should explore integration with multi-agent systems to enable coordinated analysis across domains and decision processes, which is essential for realistic climate decision support.

\section{Methods}

\subsection{Climate Data and Benchmark}

\hypertarget{climate-corpus-construction}{%
\subsubsection{\texorpdfstring{\emph{Climate Corpus Construction}}{Climate Corpus Construction}}\label{climate-corpus-construction}}

We construct a large-scale climate corpus comprising 78 billion tokens to expose the model to multidisciplinary climate knowledge spanning physical, ecological, economic and policy domains. In the absence of high-quality open-source corpora tailored for climate language model training, the dataset is assembled by systematically integrating heterogeneous sources, including assessment reports, policy documents, academic literature, books, patents and web-based materials.

To reflect varying degrees of climate relevance, we incorporate materials ranging from highly climate-specific sources to more general texts. Climate assessment reports and policy documents from international organizations, including the IPCC, UNFCCC, IEA and WMO, are retained in full to preserve authoritative knowledge (see Supplement~B1). In addition, climate policy texts are incorporated from the Global Climate Change Mitigation Policy Dataset (GCCMPD)\numcite{27}, which provides large-scale coverage of mitigation and adaptation policies across countries and regions. Academic publications and patents are filtered using climate-relevant terms identified from climate-specific texts via term frequency--inverse document frequency (TF--IDF)\numcite{28}.

Large-scale web data are further refined using a lightweight neural classifier\numcite{29} trained on curated climate data to identify climate-relevant content (see Supplement~B2). All sources are combined with graded deduplication and quality filtering to produce a unified corpus for training. The token distribution across data sources is reported in Supplement~B3.

\hypertarget{climainstruct-expert-annotated-instruction-design}{%
\subsubsection{\texorpdfstring{\emph{ClimaInstruct: Expert-Annotated Instruction Design}}{ClimaInstruct: Expert-Annotated Instruction Design}}\label{climainstruct-expert-annotated-instruction-design}}

To align the model with expert-level climate reasoning, we develop ClimaInstruct, a multidisciplinary instruction dataset annotated by 114 climate scientists from diverse fields. Rather than focusing solely on factual correctness, ClimaInstruct is designed to capture how experts formulate questions, integrate knowledge across domains and construct structured analyses.

The dataset is constructed through a pipeline combining large-scale candidate generation and expert-driven refinement. An initial pool of candidate question--answer pairs is derived from IPCC assessment reports, which provide authoritative and comprehensive coverage of climate science across physical processes, impacts, adaptation and mitigation. These materials are transformed into a diverse set of climate-relevant instructions, yielding approximately 72,000 candidates\numcite{30} (see Supplementary Section C1).

Expert annotation is then conducted to ensure both scientific validity and analytical quality. Each candidate is evaluated in terms of relevance, complexity and completeness. Here, complexity refers to the extent to which a question requires integrating concepts, evidence or mechanisms across domains, while completeness assesses whether the answer adequately addresses all aspects of the question. In addition to scoring, experts revise or rewrite responses when necessary to ensure clarity, correctness and coherence. Each sample is independently reviewed by multiple experts to reduce subjectivity. After expert filtering, 42,175 instructions are retained, and selecting samples with high ratings in both complexity and completeness yields a final set of 19,110 high-quality instructions (see Supplements~C2 and C3).

The resulting dataset encodes interdisciplinary reasoning patterns in climate science. Disciplinary labels are assigned \numcite{31} and verified to characterize cross-domain coverage (see Supplement~C4).

\hypertarget{climabench-benchmark-for-interdisciplinary-climate-reasoning}{%
\subsubsection{\texorpdfstring{\emph{ClimaBench: Benchmark for Interdisciplinary Climate Reasoning}}{ClimaBench: Benchmark for Interdisciplinary Climate Reasoning}}\label{climabench-benchmark-for-interdisciplinary-climate-reasoning}}

We develop ClimaBench to evaluate climate language models across both general climate understanding and specialized climate reasoning. The benchmark is constructed from multiple sources, including expert contributions, university-level climate science textbooks, webpages of international climate organizations and publicly available question--answer datasets. After cleaning and validation, tasks are organized into two difficulty tiers: general-level tasks, which assess foundational and commonsense climate knowledge, and specialized-level tasks, which target professional-level understanding.

To capture different forms of climate inquiry, the benchmark includes five task formats: true/false, single-choice, multiple-choice, open-ended and reasoning questions (see Supplements~D1 and D3). Reasoning questions are selected by climate domain experts based on solving them requires explicit logical inference or quantitative reasoning beyond direct retrieval of information. For these tasks, both reasoning processes and final answers are evaluated separately, enabling distinction between the correctness of conclusions and the quality of intermediate reasoning.

To support fine-grained analysis of interdisciplinary performance, each task is annotated using a standardized disciplinary taxonomy, with manual verification to ensure consistency in multi-label assignments\numcite{31}. The resulting benchmark covers 19 climate-relevant disciplines, with detailed distributions reported in Supplementary Section D.

\subsection{Training an LLM for Climate Science}

\hypertarget{alignment-for-climate-reasoning}{%
\subsubsection{\texorpdfstring{\emph{Alignment for Climate Reasoning}}{Alignment for Climate Reasoning}}\label{alignment-for-climate-reasoning}}

To align a general-purpose language model with climate science, we adopt a training strategy that combines continued pre-training (CPT)\numcite{32,33} and supervised fine-tuning (SFT)\numcite{34,35}, using Qwen2.5-32B\numcite{36} as the base model. This approach is designed to incorporate domain-specific knowledge while preserving general reasoning capacity.

Continued pre-training exposes the model to the constructed climate corpus, which spans physical climate processes, environmental systems, energy and engineering, economics and social sciences. To mitigate over-specialization and maintain general reasoning ability, the training data are complemented with non-climate corpora, mathematics, code and encyclopedic texts. This combination enables the model to integrate climate-relevant knowledge within a broader framework of logical and symbolic reasoning. The training process was implemented using the open-source LlamaFactory framework\numcite{37}, in combination with DeepSpeed for distributed training\numcite{38} (see Supplement~E1).

Supervised fine-tuning further aligns the model with expert reasoning patterns using ClimaInstruct. These instructions reflect how climate scientists formulate questions, integrate evidence across domains and construct structured analyses. The fine-tuning process was implemented with the LlamaFactory framework, incorporating a set of optimization techniques such as Flash attention\numcite{39}, S\textsuperscript{2} attention\numcite{40}, and Mixed precision\numcite{41}. Then we employed model parameter averaging techniques\numcite{42} to further stabilize training dynamics and enhance generalization performance across downstream climate science tasks. Through instruction tuning, the model learns to generate responses that not only incorporate relevant knowledge, but also follow coherent reasoning processes consistent with expert practice. To balance domain specialization and generalization, instruction data include both climate-specific and general-purpose samples (see Supplement~E2). CPT and SFT enable the model to integrate multidisciplinary climate knowledge with structured reasoning capability, forming the basis for consistent analytical performance in complex climate tasks.

\hypertarget{training-validation}{%
\subsubsection{\texorpdfstring{\emph{Training Validation}}{Training Validation}}\label{training-validation}}

To assess the effectiveness of the training strategy, we conduct validation experiments to isolate the contributions of different components in the alignment process. Specifically, we compare four model variants: the base model, models trained with CPT only, SFT only, and the combined CPT+SFT model (CFM). These comparisons are performed on ClimaBench to evaluate how domain-specific knowledge and instruction alignment jointly contribute to performance improvements in climate tasks.

To further examine generalization, we evaluate the models on additional benchmarks with varying degrees of relevance to climate science. Pira 2.0\numcite{43} and CDP-QA\numcite{44} are included to assess performance on climate- and environment-related question answering, while CEval\numcite{45} is used to evaluate general reasoning and knowledge capabilities. In addition to the base model, we include ClimateGPT-70B as a domain-specific baseline for comparison.

All evaluations are conducted using zero-shot prompting to ensure a consistent and comparable assessment of model capabilities without task-specific tuning (see Supplements~E4 and G1--G5).

\subsection{Evaluation Design}

\hypertarget{climabench-evaluation-of-climate-knowledge-and-scientific-reasoning}{%
\subsubsection{\texorpdfstring{\emph{ClimaBench Evaluation of Climate Knowledge and Scientific Reasoning}}{ClimaBench Evaluation of Climate Knowledge and Scientific Reasoning}}\label{climabench-evaluation-of-climate-knowledge-and-scientific-reasoning}}

We use ClimaBench to evaluate whether the model is aligned with established climate knowledge and can apply this knowledge across different forms of scientific inquiry. All evaluations are conducted in zero-shot settings to assess how well climate-relevant knowledge and reasoning can be transferred without task-specific adaptation.

\textbf{Closed-form tasks.} True/false, single-choice and multiple-choice questions are used to examine whether the model can correctly retrieve and apply climate-relevant knowledge when answers are explicitly defined. For true/false and single-choice questions, performance is measured by exact-match accuracy. For multiple-choice questions, agreement with the reference answer is quantified using Jaccard similarity\numcite{46}, which captures both omission of correct options and selection of incorrect ones.

\textbf{Open-form scientific reasoning.} Because these tasks do not admit reliable exact-match evaluation, we evaluate open-form responses using standardized scoring schemes\numcite{47,48}. Open-form objective questions are scored on a 4-scale Likert scoring scheme inspired by Zhou et al.\numcite{49}, where higher scores indicate greater accuracy, completeness and alignment with the question intent. Reasoning questions are evaluated on a three-point scale, with reasoning process and reasoning outcome assessed separately to distinguish between partial and fully consistent reasoning. Detailed evaluation prompts and scoring criteria are provided in the Supplementary Information, Supplements~G4 and G5.

\textbf{Interdisciplinary stress test.} To examine whether alignment persists under increasing interdisciplinary complexity, we group tasks by disciplinary breadth, defined as the number of disciplinary labels associated with each task. This stress test evaluates whether model performance remains stable as climate questions require coordination across a broader range of domains.

\hypertarget{expert-designed-open-ended-climate-problems}{%
\subsubsection{\texorpdfstring{\emph{Expert-Designed Open-Ended Climate Problems}}{Expert-Designed Open-Ended Climate Problems}}\label{expert-designed-open-ended-climate-problems}}

To evaluate whether models can support realistic climate analysis beyond structured tasks, we construct a set of 220 expert-designed open-ended climate problems. These questions are formulated by climate scientists to reflect real-world analytical scenarios across seven thematic domains, including mitigation and energy transitions, adaptation and climate risk management, climate--health--food systems, carbon removal and accounting, governance and political economy, climate system dynamics, and equity and societal transformation. Addressing these problems requires integrating knowledge spanning physical processes, socio-economic dynamics and policy considerations. Unlike benchmark-style tasks, these problems do not admit a single canonical answer, but instead emphasize professional judgment, cross-disciplinary synthesis and structured reasoning. To enable comprehensive comparison, we evaluate a broader set of state-of-the-art language models in addition to the base model and domain-specific baselines.

\textbf{Knowledge coverage and alignment.} We first analyze the breadth and alignment of knowledge expressed in model responses. For each response, disciplinary breadth is quantified by annotating the set of climate-relevant disciplines involved, allowing assessment of cross-domain coverage. To further examine alignment with established scientific knowledge, we compute semantic similarity between model-generated responses and IPCC assessment reports using an embedding-based retrieval framework\numcite{50}, providing a graded measure of proximity to authoritative climate knowledge (see Supplement~H4).

\textbf{Analytical structure and breakdown.} We further assess whether model responses follow a coherent analytical structure when addressing open-ended climate problems. For each thematic domain, we define an ordered set of expected analytical components derived from expert knowledge to reflect how climate scientists typically structure their analysis. Responses are then evaluated according to the proportion of these components that are substantively addressed. In addition, we identify the earliest point at which the analytical process breaks down, providing a structural diagnostic of where reasoning ceases to follow the expected pathway.

\textbf{Trade-off analysis.} We then evaluate whether model responses explicitly recognize and analyze trade-offs between competing objectives. Climate problems inherently involve balancing multiple goals, such as mitigation effectiveness, economic feasibility and social equity\numcite{51,52}. Responses are scored on a three-level scale: 0 if no trade-offs are identified, 1 if trade-offs are mentioned without further analysis, and 2 if alternative pathways are explicitly compared and their implications are discussed. This criterion assesses whether the model can move beyond single-solution narratives to structured comparison of competing strategies.

\textbf{Uncertainty handling.} Finally, we assess whether model responses incorporate uncertainty into the analysis of climate transition pathways. In climate transition assessment, uncertainty may arise from multiple sources\numcite{53,54}, including future climate trajectories, technological development, policy implementation and other evolving boundary conditions. Responses are scored on a three-level scale: 0 for deterministic analyses that ignore uncertainty, 1 for acknowledging uncertainty without integrating it into the analysis, and 2 for explicitly incorporating uncertainty into the analytical framework, for example through conditional pathways, scenario-dependent reasoning or robust strategies. This dimension evaluates whether the model can reason about climate transitions under uncertainty.

\section*{References}
\begin{enumerate}
\setlength{\itemsep}{2pt}
\small
\item Price, I. \emph{et al.} Probabilistic weather forecasting with machine learning. \emph{Nature} \textbf{637}, 84--90 (2025).
\item Terhaar, J., Burger, F. A., Vogt, L., Frölicher, T. L. \& Stocker, T. F. Record sea surface temperature jump in 2023--2024 unlikely but not unexpected. \emph{Nature} 1--5 (2025) doi:10.1038/s41586-025-08674-z.
\item Huang, T., Zanocco, C., Wang, Z., Hwang, J. \& Rajagopal, R. Built environment disparities are amplified during extreme weather recovery. \emph{Nature} \textbf{648}, 349--356 (2025).
\item Fanning, A. L. \& Raworth, K. Doughnut of social and planetary boundaries monitors a world out of balance. \emph{Nature} \textbf{646}, 47--56 (2025).
\item Stechemesser, A. \emph{et al.} Climate policies that achieved major emission reductions: Global evidence from two decades. \emph{Science} \textbf{385}, 884--892 (2024).
\item Wu, L., Liu, G., Huang, Z., Meng, J. \& Zhou, Y. Cross-national comparative assessment of synergies and conflicts in climate policy mixes. \emph{Nat. Clim. Change} 1--10 (2026) doi:10.1038/s41558-026-02574-4.
\item Shorey, P. \& Abdulla, A. Integrating climate and physical constraints into assessments of net capture from direct air capture facilities. \emph{Proc. Natl. Acad. Sci.} \textbf{122}, e2410824121 (2025).
\item van Vuuren, D. P. \emph{et al.} Exploring pathways for world development within planetary boundaries. \emph{Nature} \textbf{641}, 910--916 (2025).
\item IPCC --- Intergovernmental Panel on Climate Change. https://www.ipcc.ch/.
\item Sognnaes, I. \& Peters, G. P. Influence of individual models and studies on quantitative mitigation findings in the IPCC Sixth Assessment Report. \emph{Nat. Commun.} \textbf{16}, 8343 (2025).
\item van de Ven, D. J. \emph{et al.} Energy and socioeconomic system transformation through a decade of IPCC-assessed scenarios. \emph{Nat. Clim. Change} \textbf{15}, 218--226 (2025).
\item COP29: involve the IPCC in defining climate finance. \emph{Nature} \textbf{635}, 7--8 (2024).
\item Bodin, S. \& Gustafsson, Ö. Improving the IPCC--UNFCCC relationship for effective provision of policy-relevant science. \emph{Nat. Clim. Change} \textbf{15}, 910--911 (2025).
\item Vaghefi, S. A. \emph{et al.} ChatClimate: Grounding conversational AI in climate science. \emph{Commun. Earth Environ.} \textbf{4}, 1--13 (2023).
\item Sietsma, A. J., Ford, J. D. \& Minx, J. C. The next generation of machine learning for tracking adaptation texts. \emph{Nat. Clim. Change} \textbf{14}, 31--39 (2024).
\item Chang, C. H. \emph{et al.} Global evidence of human well-being and biodiversity impacts of natural climate solutions. \emph{Nat. Sustain.} \textbf{8}, 75--85 (2025).
\item Hicks, C., Davidson, K., Lau, J. H. \& Nguyen, T. M. P. Implications of declaration of climate emergency on Australian local government policy in the State of Victoria: policy analysis utilising an LLM-based retriever-reader pipeline. \emph{Clim. Change} \textbf{178}, 191 (2025).
\item Wu, Y. \emph{et al.} CLEAR: Climate Policy Retrieval and Summarization Using LLMs. in \emph{Companion Proceedings of the ACM on Web Conference 2025} 2927--2930 (Association for Computing Machinery, New York, NY, USA, 2025). doi:10.1145/3701716.3715170.
\item Cao, C., Zhuang, J. \& He, Q. LLM-Assisted Modeling and Simulations for Public Sector Decision-Making: Bridging Climate Data and Policy Insights. in (2024).
\item Koldunov, N. \& Jung, T. Local climate services for all, courtesy of large language models. \emph{Commun. Earth Environ.} \textbf{5}, 13 (2024).
\item Pan, H., Adamu, M., Zhang, Q., Dragut, E. \& Latecki, L. J. ClimateIE: A Dataset for Climate Science Information Extraction. in \emph{Proceedings of the 2nd Workshop on Natural Language Processing Meets Climate Change (ClimateNLP 2025)} (eds Dutia, K. et al.) 76--98 (Association for Computational Linguistics, Vienna, Austria, 2025). doi:10.18653/v1/2025.climatenlp-1.6.
\item Atkins, C., Girgente, G., Shirzaei, M. \& Kim, J. Generative AI tools can enhance climate literacy but must be checked for biases and inaccuracies. \emph{Commun. Earth Environ.} \textbf{5}, 226 (2024).
\item Geiger, A. \emph{et al.} Causal Abstraction: A Theoretical Foundation for Mechanistic Interpretability. Preprint at https://doi.org/10.48550/arXiv.2301.04709 (2023).
\item Jin, Z. \emph{et al.} Can Large Language Models Infer Causation from Correlation? in (2024).
\item Plaat, A. \emph{et al.} Multi-Step Reasoning with Large Language Models, a Survey. \emph{ACM Comput Surv} \textbf{58}, 160:1-160:35 (2025).
\item Nguyen, V., Karimi, S., Hallgren, W. \& Prakash, M. Question Answering in Climate Adaptation for Agriculture: Model Development and Evaluation with Expert Feedback. in \emph{Findings of the Association for Computational Linguistics: ACL 2025} (eds Che, W., Nabende, J., Shutova, E. \& Pilehvar, M. T.) 7045--7075 (Association for Computational Linguistics, Vienna, Austria, 2025). doi:10.18653/v1/2025.findings-acl.368.
\item Wu, L., Huang, Z., Zhang, X. \& Wang, Y. Harmonizing existing climate change mitigation policy datasets with a hybrid machine learning approach. \emph{Sci. Data} \textbf{11}, 580 (2024).
\item Sparck Jones, K. A Statistical Interpretation of Term Specificity and Its Application in Retrieval. \emph{J. Doc.} \textbf{28}, 11--21 (1972).
\item Joulin, A. \emph{et al.} FastText.zip: Compressing text classification models. Preprint at https://arxiv.org/abs/1612.03651v1 (2016).
\item Wan, Y. \emph{et al.} SciQAG: A Framework for Auto-Generated Science Question Answering Dataset with Fine-grained Evaluation. Preprint at https://doi.org/10.48550/arXiv.2405.09939 (2024).
\item Pangakis, N. \& Wolken, S. Knowledge Distillation in Automated Annotation: Supervised Text Classification with LLM-Generated Training Labels. in \emph{Proceedings of the Sixth Workshop on Natural Language Processing and Computational Social Science (NLP+CSS 2024)} (eds Card, D., Field, A., Hovy, D. \& Keith, K.) 113--131 (Association for Computational Linguistics, Mexico City, Mexico, 2024). doi:10.18653/v1/2024.nlpcss-1.9.
\item Uppaal, R., Li, Y. \& Hu, J. How Useful is Continued Pre-Training for Generative Unsupervised Domain Adaptation? in \emph{Proceedings of the 9th Workshop on Representation Learning for NLP (RepL4NLP-2024)} (eds Zhao, C. et al.) 99--117 (Association for Computational Linguistics, Bangkok, Thailand, 2024).
\item Chen, J. \emph{et al.} Towards Effective and Efficient Continual Pre-training of Large Language Models. in \emph{Proceedings of the 63rd Annual Meeting of the Association for Computational Linguistics (Volume 1: Long Papers)} (eds Che, W., Nabende, J., Shutova, E. \& Pilehvar, M. T.) 5779--5795 (Association for Computational Linguistics, Vienna, Austria, 2025). doi:10.18653/v1/2025.acl-long.289.
\item Dong, G. \emph{et al.} How Abilities in Large Language Models are Affected by Supervised Fine-tuning Data Composition. in \emph{Proceedings of the 62nd Annual Meeting of the Association for Computational Linguistics (Volume 1: Long Papers)} (eds Ku, L.-W., Martins, A. \& Srikumar, V.) 177--198 (Association for Computational Linguistics, Bangkok, Thailand, 2024). doi:10.18653/v1/2024.acl-long.12.
\item Ye, J. \emph{et al.} Analyzing the Effects of Supervised Fine-Tuning on Model Knowledge from Token and Parameter Levels. in \emph{Proceedings of the 2025 Conference on Empirical Methods in Natural Language Processing} (eds Christodoulopoulos, C., Chakraborty, T., Rose, C. \& Peng, V.) 471--513 (Association for Computational Linguistics, Suzhou, China, 2025). doi:10.18653/v1/2025.emnlp-main.25.
\item Qwen \emph{et al.} Qwen2.5 Technical Report. Preprint at https://doi.org/10.48550/arXiv.2412.15115 (2025).
\item Zheng, Y., Zhang, R., Zhang, J., Ye, Y. \& Luo, Z. LlamaFactory: Unified Efficient Fine-Tuning of 100+ Language Models. in \emph{Proceedings of the 62nd Annual Meeting of the Association for Computational Linguistics (Volume 3: System Demonstrations)} (eds Cao, Y., Feng, Y. \& Xiong, D.) 400--410 (Association for Computational Linguistics, Bangkok, Thailand, 2024). doi:10.18653/v1/2024.acl-demos.38.
\item Rajbhandari, S., Rasley, J., Ruwase, O. \& He, Y. ZeRO: Memory optimizations Toward Training Trillion Parameter Models. in \emph{SC20: International Conference for High Performance Computing, Networking, Storage and Analysis} 1--16 (2020). doi:10.1109/SC41405.2020.00024.
\item Shah, J. \emph{et al.} FlashAttention-3: Fast and Accurate Attention with Asynchrony and Low-precision. \emph{Adv. Neural Inf. Process. Syst.} \textbf{37}, 68658--68685 (2024).
\item Lin, X. \emph{et al.} S2-ATTENTION: HARDWARE-AWARE CONTEXT SHARDING AMONG ATTENTION HEADS. in \emph{Proceedings of the International Conference on Learning Representations (ICLR)} (2025).
\item Dettmers, T., Lewis, M., Belkada, Y. \& Zettlemoyer, L. GPT3.int8(): 8-bit Matrix Multiplication for Transformers at Scale. \emph{Adv. Neural Inf. Process. Syst.} \textbf{35}, 30318--30332 (2022).
\item Shi, S. \emph{et al.} ULMR: Unlearning Large Language Models via Negative Response and Model Parameter Average. in \emph{Proceedings of the 2024 Conference on Empirical Methods in Natural Language Processing: Industry Track} (eds Dernoncourt, F., Preoţiuc-Pietro, D. \& Shimorina, A.) 755--762 (Association for Computational Linguistics, Miami, Florida, US, 2024). doi:10.18653/v1/2024.emnlp-industry.57.
\item Paschoal, A. F. A. \emph{et al.} Pirá: A Bilingual Portuguese-English Dataset for Question-Answering about the Ocean. in \emph{Proceedings of the 30th ACM International Conference on Information \& Knowledge Management} 4544--4553 (Association for Computing Machinery, New York, NY, USA, 2021). doi:10.1145/3459637.3482012.
\item Spokoyny, D., Laud, T., Corringham, T. \& Berg-Kirkpatrick, T. Towards Answering Climate Questionnaires from Unstructured Climate Reports. Preprint at https://doi.org/10.48550/arXiv.2301.04253 (2023).
\item Nguyen, V. B., Seifert, C. \& Schlötterer, J. CEval: A Benchmark for Evaluating Counterfactual Text Generation. in \emph{Proceedings of the 17th International Natural Language Generation Conference} (eds Mahamood, S., Minh, N. L. \& Ippolito, D.) 55--69 (Association for Computational Linguistics, Tokyo, Japan, 2024).
\item Shin, D., Lee, S., Kovacec, K. L. \& Kim, S. From Generation to Selection: Findings of Converting Analogical Problem-Solving into Multiple-Choice Questions. in \emph{Findings of the Association for Computational Linguistics: EMNLP 2024} (eds Al-Onaizan, Y., Bansal, M. \& Chen, Y.-N.) 6696--6708 (Association for Computational Linguistics, Miami, Florida, USA, 2024). doi:10.18653/v1/2024.findings-emnlp.392.
\item Singhal, K. \emph{et al.} Large language models encode clinical knowledge. \emph{Nature} \textbf{620}, 172--180 (2023).
\item Wang, C. \emph{et al.} Evaluating Open-QA Evaluation. \emph{Adv. Neural Inf. Process. Syst.} \textbf{36}, 77013--77042 (2023).
\item Zhou, C. \emph{et al.} LIMA: Less Is More for Alignment. \emph{Adv. Neural Inf. Process. Syst.} \textbf{36}, 55006--55021 (2023).
\item Vector embeddings \textbar{} OpenAI API. \url{https://developers.openai.com/api/docs/guides/embeddings/}.
\item Bauer, N. \emph{et al.} Quantification of an efficiency--sovereignty trade-off in climate policy. \emph{Nature} \textbf{588}, 261--266 (2020).
\item Prather, M. J., Gettelman, A. \& Penner, J. E. Trade-offs in aviation impacts on climate favour non-CO2 mitigation. \emph{Nature} \textbf{643}, 988--993 (2025).
\item Bevacqua, E., Fischer, E., Sillmann, J. \& Zscheischler, J. Moderate global warming does not rule out extreme global climate outcomes. \emph{Nature} \textbf{651}, 946--953 (2026).
\item Burke, M., Zahid, M., Diffenbaugh, N. S. \& Hsiang, S. Quantifying climate loss and damage consistent with a social cost of carbon. \emph{Nature} \textbf{651}, 959--966 (2026).
\item Wallach, H. M. Topic modeling: beyond bag-of-words. in \emph{Proceedings of the 23rd international conference on Machine learning} 977--984 (Association for Computing Machinery, New York, NY, USA, 2006). doi:10.1145/1143844.1143967.
\item Peng, H., Li, J., Song, Y. \& Liu, Y. Incrementally Learning the Hierarchical Softmax Function for Neural Language Models. \emph{Proceedings of the AAAI Conference on Artificial Intelligence} \textbf{31}, (2017).
\item Paster, K., Santos, M. D., Azerbayev, Z. \& Ba, J. OpenWebMath: An Open Dataset of High-Quality Mathematical Web Text. Preprint at https://doi.org/10.48550/arXiv.2310.06786 (2023).
\item Azerbayev, Z. \emph{et al.} Llemma: An Open Language Model For Mathematics. Preprint at https://doi.org/10.48550/arXiv.2310.10631 (2024).
\item Yue, X. \emph{et al.} MAmmoTH: Building Math Generalist Models through Hybrid Instruction Tuning. Preprint at https://doi.org/10.48550/arXiv.2309.05653 (2023).
\item Li, K. \emph{et al.} InstructCoder: Instruction Tuning Large Language Models for Code Editing. Preprint at https://doi.org/10.48550/arXiv.2310.20329 (2024).
\item Penedo, G. \emph{et al.} The FineWeb Datasets: Decanting the Web for the Finest Text Data at Scale. \emph{Advances in Neural Information Processing Systems} \textbf{37}, 30811--30849 (2024).
\item Guo, M., Dai, Z., Vrandečić, D. \& Al-Rfou, R. Wiki-40B: Multilingual Language Model Dataset. in \emph{Proceedings of the Twelfth Language Resources and Evaluation Conference} (eds Calzolari, N. et al.) 2440--2452 (European Language Resources Association, Marseille, France, 2020).
\item Hayou, S., Ghosh, N. \& Yu, B. LoRA+: Efficient Low Rank Adaptation of Large Models. Preprint at https://doi.org/10.48550/arXiv.2402.12354 (2024).
\item Min, S. \emph{et al.} FActScore: Fine-grained Atomic Evaluation of Factual Precision in Long Form Text Generation. in \emph{Proceedings of the 2023 Conference on Empirical Methods in Natural Language Processing} (eds Bouamor, H., Pino, J. \& Bali, K.) 12076--12100 (Association for Computational Linguistics, Singapore, 2023). doi:10.18653/v1/2023.emnlp-main.741.
\item Smith, C. P. \emph{Motivation and Personality: Handbook of Thematic Content Analysis}. (Cambridge University Press, 1992).
\item Liu, Y. \emph{et al.} G-Eval: NLG Evaluation using Gpt-4 with Better Human Alignment. in \emph{Proceedings of the 2023 Conference on Empirical Methods in Natural Language Processing} (eds Bouamor, H., Pino, J. \& Bali, K.) 2511--2522 (Association for Computational Linguistics, Singapore, 2023). doi:10.18653/v1/2023.emnlp-main.153.
\item Zheng, L. \emph{et al.} Judging LLM-as-a-Judge with MT-Bench and Chatbot Arena. \emph{Advances in Neural Information Processing Systems} \textbf{36}, 46595--46623 (2023).
\item text-embedding-3-large Model \textbar{} OpenAI API. \url{https://developers.openai.com/api/docs/models/text-embedding-3-large}.
\item Yao, S. \emph{et al.} REACT: SYNERGIZING REASONING AND ACTING IN LANGUAGE MODELS. in (2023).
\item Wei, J. \emph{et al.} Chain-of-Thought Prompting Elicits Reasoning in Large Language Models. \emph{Advances in Neural Information Processing Systems} \textbf{35}, 24824--24837 (2022).
\item Li, P. \emph{et al.} Using deep learning to generate key variables in global mitigation scenarios. \emph{Nat. Clim. Chang.} \textbf{15}, 760--768 (2025).
\item Qin, Y. \emph{et al.} ToolLLM: Facilitating Large Language Models to Master 16000+ Real-world APIs. in \emph{12th International Conference on Learning Representations} (2023).
\end{enumerate}
\clearpage
\onecolumn
\newgeometry{top=1.7cm,bottom=1.7cm,left=1.5cm,right=1.5cm}
\appendix
\renewcommand{\thesubsection}{\thesection\arabic{subsection}}
\footnotesize
\setlength{\tabcolsep}{3pt}
\sloppy
\section*{Appendix: Supplementary Information}
\addcontentsline{toc}{section}{Appendix: Supplementary Information}
\addtocontents{toc}{\protect\setcounter{tocdepth}{2}}
\tableofcontents
\clearpage
\begin{multicols}{2}
\section{Disciplines in Climate Science}

\subsection{Category and abbreviation of climate-related disciplines}

All instructions in ClimaInstruct and tasks in ClimaBench are labeled based on the Scopus ASJC classification scheme. Rather than adopting the full ASJC taxonomy, we selectively identify a subset of climate-relevant disciplines, covering 19 tier 1 subject and 144 fine-grained tier 2 subject across the physical sciences, social sciences, and life and health sciences.

Table A1 presents the selected climate-related disciplines and their corresponding abbreviations used throughout this paper.

\begin{center}
\textbf{Table A1. Selected climate-related disciplines from Scopus ASJC}
\end{center}

\begingroup\scriptsize\setlength{\tabcolsep}{2pt}
\begin{tabularx}{\linewidth}{@{}lXl@{}}
\toprule
\textbf{Field} & \textbf{Tier 1 Subject} & \textbf{Abb.} \\
\midrule
Physical Sciences & Chemical Engineering & ChE \\
Physical Sciences & Chemistry & Chem \\
Physical Sciences & Computer Science & CS \\
Physical Sciences & Earth and Planetary Sciences & EPS \\
Physical Sciences & Energy & Energy \\
Physical Sciences & Engineering & Eng \\
Physical Sciences & Environmental Science & EnvSci \\
Physical Sciences & Materials Science & MatSci \\
Physical Sciences & Mathematics & Math \\
Physical Sciences & Physics and Astronomy & PhysAstro \\
Social Sciences & Arts and Humanities & ArtsHum \\
Social Sciences & Business, Management and Accounting & BMA \\
Social Sciences & Decision Sciences & DecSci \\
Social Sciences & Economics, Econometrics and Finance & Econ \\
Social Sciences & Psychology & Psych \\
Social Sciences & Social Sciences & SocSci \\
Life \& Health Sciences & Agricultural and Biological Sciences & AgBio \\
Life \& Health Sciences & Medicine & Med \\
Life \& Health Sciences & Pharmacology, Toxicology and Pharmaceutics & PTP \\
\bottomrule
\end{tabularx}
\endgroup

\subsection{Hierarchical subject classifications across disciplines}

Table A2-A4 shows the tier 1 and tier 2 subject of physical sciences, social sciences, and life and health sciences.

\end{multicols}
\begin{center}
\textbf{Table A2. Tier 1 and tier 2 subject of physical sciences}
\end{center}

\begingroup\tiny\renewcommand{\arraystretch}{.70}\setlength{\tabcolsep}{2pt}
\begin{tabularx}{\textwidth}{@{}p{.13\textwidth}Xp{.13\textwidth}X@{}}
\toprule
\textbf{Tier 1 Subject} & \textbf{Tier 2 Subject} & \textbf{Tier 1 Subject} & \textbf{Tier 2 Subject} \\
\midrule
Chemical Engineering & General Chemical Engineering & Engineering & General Engineering \\
 & Chemical Engineering (miscellaneous) &  & Aerospace Engineering \\
 & Bioengineering &  & Automotive Engineering \\
 & Catalysis &  & Biomedical Engineering \\
 & Chemical Health and Safety &  & Electrical and Electronic Engineering \\
 & Colloid and Surface Chemistry &  & Industrial and Manufacturing Engineering \\
 & Filtration and Separation &  & Mechanical Engineering \\
 & Fluid Flow and Transfer Processes &  & Mechanics of Materials \\
 & Process Chemistry and Technology &  & Ocean Engineering \\
Chemistry & General Chemical Engineering &  & Building and Construction \\
 & Bioengineering & Environmental Science & General Environmental Science \\
 & General Chemistry &  & Ecological Modelling \\
 & Analytical Chemistry &  & Ecology \\
 & Electrochemistry &  & Environmental Chemistry \\
 & Inorganic Chemistry &  & Environmental Engineering \\
 & Organic Chemistry &  & Global and Planetary Change \\
 & Spectroscopy &  & Health, Toxicology and Mutagenesis \\
Computer Science & General Computer Science &  & Management, Monitoring, Policy and Law \\
 & Artificial Intelligence &  & Nature and Landscape Conservation \\
 & Computer Vision and Pattern Recognition &  & Pollution \\
 & Information Systems &  & Waste Management and Disposal \\
 & Signal Processing &  & Water Science and Technology \\
Earth and Planetary Sciences & General Earth and Planetary Sciences & Materials Science & General Materials Science \\
 & Atmospheric Science &  & Biomaterials \\
 & Computers in Earth Sciences &  & Ceramics and Composites \\
 & Earth-Surface Processes &  & Electronic, Optical and Magnetic Materials \\
 & Economic Geology &  & Materials Chemistry \\
 & Geochemistry and Petrology &  & Metals and Alloys \\
 & Geology &  & Polymers and Plastics \\
 & Geophysics &  & Surfaces, Coatings and Films \\
 & Geotechnical Engineering and Engineering Geology & Mathematics & General Mathematics \\
 & Oceanography &  & Applied Mathematics \\
 & Palaeontology &  & Modelling and Simulation \\
 & Space and Planetary Science &  & Numerical Analysis \\
 & Stratigraphy &  & Statistics and Probability \\
Energy & General Energy &  & Theoretical Computer Science \\
 & Energy Engineering and Power Technology & Physics and Astronomy & General Physics and Astronomy \\
 & Fuel Technology &  & Physics and Astronomy (miscellaneous) \\
 & Nuclear Energy and Engineering &  & Acoustics and Ultrasonics \\
 & Renewable Energy, Sustainability and the Environment &  & Astronomy and Astrophysics \\
 &  &  & Condensed Matter Physics \\
 &  &  & Instrumentation \\
 &  &  & Nuclear and High Energy Physics \\
 &  &  & Atomic and Molecular Physics \\
 &  &  & Radiation \\
 &  &  & Statistical and Nonlinear Physics \\
 &  &  & Surfaces and Interfaces \\
\bottomrule
\end{tabularx}
\endgroup
\begin{multicols}{2}

\end{multicols}
\begin{minipage}{\textwidth}
\begin{center}
\textbf{Table A3. Tier 1 and tier 2 subject of social sciences}
\end{center}

\begingroup\tiny\renewcommand{\arraystretch}{.78}\setlength{\tabcolsep}{2pt}
\begin{tabularx}{\textwidth}{@{}p{.16\textwidth}Xp{.16\textwidth}X@{}}
\toprule
\textbf{Tier 1 Subject} & \textbf{Tier 2 Subject} & \textbf{Tier 1 Subject} & \textbf{Tier 2 Subject} \\
\midrule
Arts and Humanities & General Arts and Humanities & Psychology & General Psychology \\
 & History and Philosophy of Science &  & Applied Psychology \\
 & Literature and Literary Theory & Social Sciences & General Social Sciences \\
 & Museology &  & Development \\
 & Philosophy &  & Education \\
Business, Management and Accounting & General Business, Management and Accounting &  & Geography, Planning and Development \\
 & Business and International Management &  & Health (social science) \\
 & Management Information Systems &  & Human Factors and Ergonomics \\
 & Management of Technology and Innovation &  & Law \\
 & Marketing &  & Sociology and Political Science \\
 & Organizational Behavior and Human Resource Management &  & Transportation \\
 & Tourism, Leisure and Hospitality Management &  & Cultural Studies \\
 & Industrial relations &  & Demography \\
Decision Sciences & General Decision Sciences &  & Gender Studies \\
 & Information Systems and Management &  & Life-span and Life-course Studies \\
 & Management Science and Operations Research &  & Political Science and International Relations \\
 & Statistics, Probability and Uncertainty &  & Public Administration \\
Economics, Econometrics and Finance & General Economics, Econometrics and Finance &  & Urban Studies \\
 & Economics and Econometrics &  &  \\
 & Finance &  &  \\
\bottomrule
\end{tabularx}
\endgroup
\end{minipage}
\begin{multicols}{2}

\end{multicols}
\begin{center}
\textbf{Table A4. Tier 1 and tier 2 subject of life and health sciences}
\end{center}
\begingroup
\scriptsize
\renewcommand{\arraystretch}{0.86}
\setlength{\tabcolsep}{2pt}
\setlength{\LTleft}{0pt plus 1fill}
\setlength{\LTright}{0pt plus 1fill}
\begin{longtable}[]{@{}ll@{}}
\toprule
\textbf{Tier 1 Subject} & \textbf{Tier 2 Subject} \\
\midrule
\endhead
Agricultural and Biological Sciences & General Agricultural and Biological Sciences \\
& Agronomy and Crop Science \\
& Animal Science and Zoology \\
& Aquatic Science \\
& Ecology, Evolution, Behavior and Systematics \\
& Food Science \\
& Forestry \\
& Horticulture \\
& Insect Science \\
& Plant Science \\
& Soil Science \\
Medicine & General Medicine \\
& Epidemiology \\
& Geriatrics and Gerontology \\
& Health Informatics \\
& Health Policy \\
& Public Health, Environmental and Occupational Health \\
Pharmacology, Toxicology and Pharmaceutics & General Pharmacology, Toxicology and Pharmaceutics \\
& Toxicology \\
\bottomrule
\end{longtable}
\endgroup
\begin{multicols}{2}

\section{Construction of Interdisciplinary Climate Corpus}

\subsection{Source of reports used to construct climate corpus}

Table B1 shows the full list of 17 international authoritative organizations, which is one of the most important sources for constructing interdisciplinary climate corpus.

\end{multicols}
\begin{center}
\textbf{Table B1. International authoritative organizations}
\end{center}
\begingroup
\scriptsize
\renewcommand{\arraystretch}{0.86}
\setlength{\tabcolsep}{2pt}
\setlength{\LTleft}{0pt plus 1fill}
\setlength{\LTright}{0pt plus 1fill}
\begin{longtable}[]{@{}lll@{}}
\toprule
\textbf{Organization} & \textbf{Abb.} & \textbf{Website} \\
\midrule
\endhead
European Environment Agency & EEA & \href{https://www.eea.europa.eu/en}{\underline{https://www.eea.europa.eu/en}} \\
U.S. Energy Information Administration & EIA & \href{https://www.eia.gov/}{\underline{https://www.eia.gov/}} \\
U.S. Environmental Protection Agency & EPA & \href{https://www.epa.gov/}{\underline{https://www.epa.gov/}} \\
Green Climate Fund & GCF & \href{https://www.greenclimate.fund/}{\underline{https://www.greenclimate.fund/}} \\
Global Environment Facility & GEF & \href{https://www.thegef.org/}{\underline{https://www.thegef.org/}} \\
International Energy Agency & IEA & \href{https://www.iea.org/}{\underline{https://www.iea.org/}} \\
Intergovernmental Science-Policy Platform on Biodiversity and Ecosystem Services & IPBES & \href{https://www.ipbes.net/}{\underline{https://www.ipbes.net/}} \\
Intergovernmental Panel on Climate Change & IPCC & \href{https://www.ipcc.ch/}{\underline{https://www.ipcc.ch/}} \\
International Renewable Energy Agency & IRENA & \href{https://www.irena.org/}{\underline{https://www.irena.org/}} \\
International Union for Conservation of Nature & IUCN & \href{https://www.iucn.org/}{\underline{https://www.iucn.org/}} \\
Organization of the Petroleum Exporting Countries & OPEC & \href{https://www.opec.org/opec_web/en/}{\underline{https://www.opec.org/opec\_web/en/}} \\
United Nations Development Programme & UNDP & \href{https://www.undp.org/}{\underline{https://www.undp.org/}} \\
United Nations Environment Programme & UNEP & \href{https://www.unep.org/}{\underline{https://www.unep.org/}} \\
United Nations Framework Convention on Climate Change & UNFCCC & \href{https://unfccc.int/}{\underline{https://unfccc.int/}} \\
World Meteorological Organization & WMO & \href{https://wmo.int/}{\underline{https://wmo.int/}} \\
World Wide Fund for Nature & WWF & \href{https://www.worldwildlife.org/}{\underline{https://www.worldwildlife.org/}} \\
World Bank & WB & \href{https://www.worldbank.org/en/home}{\underline{https://www.worldbank.org/en/home}} \\
\bottomrule
\end{longtable}
\endgroup
\begin{multicols}{2}

\subsection{Framework of identify climate-relevant content from web data}

\textbf{Workflow.} We obtain general sources from web-based text and scanned documents collected through web crawling, and conduct optical character recognition (OCR) applied to digitize scanned materials. Due to the scale and heterogeneity of these data, additional refinement was required to isolate climate-relevant content. Climate assessment reports and policy documents from international organizations were used as seed data to train a compact text classifier based on the FastText framework. The trained classifier was applied to the cleaned general-source data to recall climate-relevant texts. To further improve coverage, we analyzed the distribution of source URLs in the recalled dataset, identified frequently occurring base domains, and supplemented the dataset with additional webpages from these domains. The refined general-source data were then combined with climate-specialized and mixed-specialized sources to form the final climate corpus used for continued pre-training.

\textbf{FastText framework.} FastText is an open-source tool introduced by Facebook in 2016\numcite{29}, designed to efficiently handle large-scale text classification and word embedding tasks. Its core principle involves integrating the Bag-of-Words (BoW) model\numcite{55} with word embeddings, leveraging subword embeddings and hierarchical softmax to enhance computational efficiency.

Words often possess internal morphological structures and etymological relationships, which may correlate across different terms. In FastText, each central word is represented as a collection of subwords, thereby improving the quality of word representation.

Hierarchical softmax is an optimization strategy for the traditional softmax function\numcite{56}. Instead of computing the probability distribution for all words simultaneously, it constructs a binary tree structure. The tree is built based on category frequency, with high-frequency categories positioned closer to the root node. For each category, the probability is calculated as the product of node probabilities along the path from the root to the target leaf node. This approach significantly reduces computational complexity while maintaining classification accuracy. This process can be expressed as:

\begin{equation}
P(y\mid x)=\prod_{i=1}^{L}\sigma\!\left(\mathbf{h}^{\top}\mathbf{W}_{n_i}\right)
\tag{B1}\label{eq:B1}
\end{equation}

where \(L\) denotes the path length, \(\mathbf{h}\) represents the text vector, \(\mathbf{W}_{n_{i}}\) is the weight associated with node \(n_{i}\), and \(\text{\(\sigma\)\ }\)refers to the sigmoid function. By employing hierarchical softmax and parallel computing techniques, efficient training can be achieved. Moreover, FastText is capable of adapting to multilingual environments.

\subsection{Composition of the corpus by source type}

We constructed a climate corpus comprising approximately 78 billion tokens (Fig. B1) to enrich domain-relevant knowledge. The corpus integrates assessment and decision reports from 17 international organizations such as the IPCC (Supplement B1), together with academic publications, patents, books, and climate-relevant web texts.

\begin{minipage}{\linewidth}
\centering
\includegraphics[width=.82\linewidth,height=.30\textheight,keepaspectratio]{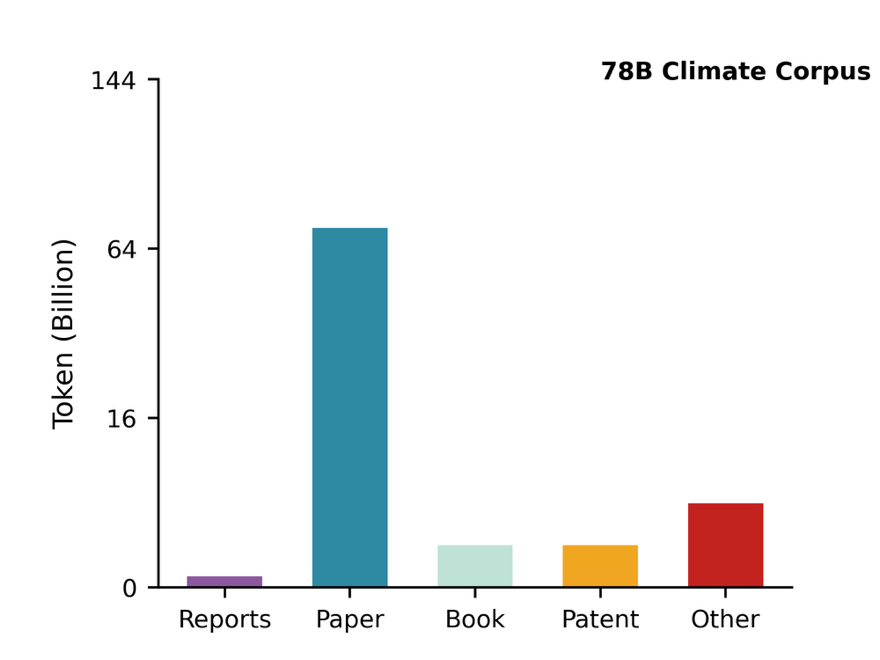}
\par\smallskip
\textbf{Figure B1. Corpus composition across data sources}
\end{minipage}

\section{ClimaInstruct}

\subsection{Pre-construction of QA dataset}

In the pre-construction stage, we selected assessment reports from the Intergovernmental Panel on Climate Change (IPCC), which also form part of the climate-specialized corpus. These reports provide authoritative and comprehensive coverage of climate science, including physical processes, impacts, adaptation, and mitigation. The reports were segmented into coherent natural language paragraphs, which were then used to generate QA pairs through prompt-based generation with GPT-o1. This process yielded an initial pool of 72,000 QA pairs.

The prompt for constructing QA dataset from the segmented reports was as follows:

\begingroup\scriptsize\setlength{\parindent}{0pt}
\par\smallskip\hrule\smallskip
System： You are an expert on climate policy.

Prompt：

The following is a paragraph related to climate science. Please raise a climate question based on the meaning of this paragraph and answer it according to the content of the paragraph.

Please note that your output language must be English. Each output should include 2 lines, namely the question and the answer.

Paragraph:\textbf{\textless{} PARAGRAPH \textgreater{}}
\par\smallskip\hrule\smallskip\endgroup

Note: The \textbf{\textless PARAGRAPH\textgreater{}} denotes the segmented and processed content extracted from the reports of the Intergovernmental Panel on Climate Change (IPCC).

\subsection{Flowchart for expert annotation}

ClimaInstruct is a climate-specific instruction dataset designed for supervised fine-tuning. It is constructed independently of the climate pretraining corpus and developed through a dedicated multi-stage pipeline that integrates expert annotation with subsequent refinement.

Figure C1 shows the flowchart for the steps of expert annotation, including the validity of the question, the correlation between question and answer, and the quality of QA pair.

\begin{minipage}{\linewidth}
\centering
\includegraphics[width=.82\linewidth,height=.34\textheight,keepaspectratio]{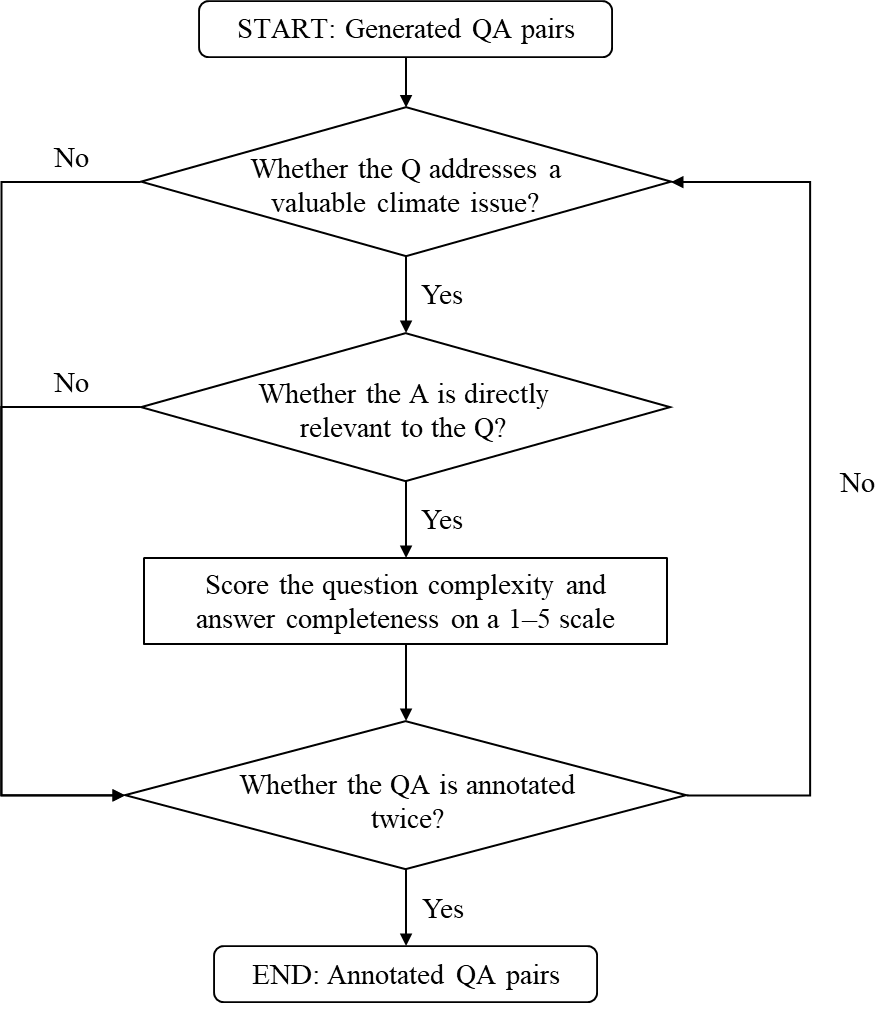}
\par\smallskip
\textbf{Figure C1. Flowchart for expert annotation}
\end{minipage}

\subsection{Filtering of ClimaInstruct}

As shown in Fig. C2, A total of 72,000 candidate question--answer (QA) pairs were initially collected and subjected to expert annotation. After this stage, 42,175 QA pairs were retained. Further filtering was applied based on quality thresholds, requiring scores of at least 4 for both complexity and completeness. This resulted in a final set of 19,110 high-quality instructions, which constitute the ClimaInstruct dataset used for supervised fine-tuning.

\begin{minipage}{\linewidth}
\centering
\includegraphics[width=.82\linewidth,height=.30\textheight,keepaspectratio]{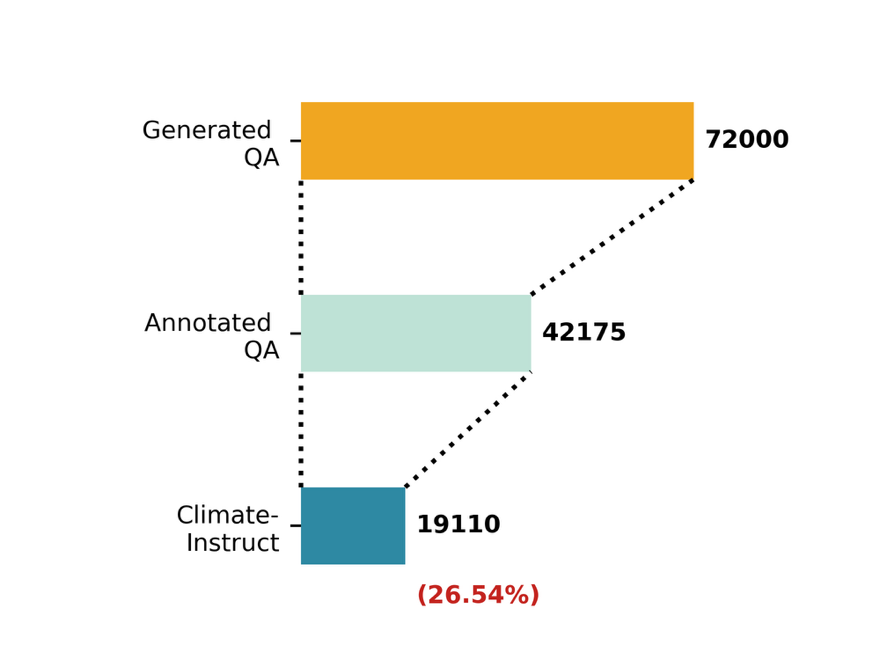}
\par\smallskip
\textbf{Figure C2. Filtering of ClimaInstruct}
\end{minipage}

\subsection{Prompt for field labeling of ClimaInstruct}

To ensure interdisciplinary coverage, all instructions in ClimaInstruct were further labeled according to the discipline standard outlined in Supplement A2. Field labeling was initially performed using GPT-o1\numcite{31}. Instructions involving multiple disciplines were assigned multiple labels where appropriate. Manual verification and correction followed to ensure high labeling fidelity.

The prompt for field labeling of Climate-Instruct was as follows:

\begingroup\scriptsize\setlength{\parindent}{0pt}
\par\smallskip\hrule\smallskip
System: You are an expert on climate policy.

Prompt:

I have a list that includes many types of subjects. Please determine which subject type the following question belongs to. You only need to answer with the subject type in the list.

If you think the question belongs to multiple subject types, please separate them with a ",".

Please answer the question in English.

If you think none of the above options apply, then answer with the closest subject type (it must be within the subject list I provided).

\textbf{\textless{} SUBJECT LIST \textgreater{}}

Question:\textbf{\textless QUESTION\textgreater{}}
\par\smallskip\hrule\smallskip\endgroup

Note: The \textbf{\textless{} SUBJECT LIST \textgreater{}} is provided in Supplement A2, and \textbf{\textless QUESTION\textgreater{}} refers to the instruction in ClimaInstruct.

\section{ClimaBench}

\subsection{Task formats and levels of domain specificity structure of ClimaBench}

As shown in Table D1, ClimaBench comprises multiple task types to evaluate both knowledge accuracy and reasoning consistency. The tasks also span both general-level and specialized-level questions. General-level tasks assess foundational climate knowledge and commonly shared concepts, whereas specialized-level tasks target professional climate expertise and require ability of discipline-specific theories, methods, and scientific reasoning.

\end{multicols}
\begin{center}
\textbf{Table D1. Statistics of task formats and specificity in ClimaBench}
\end{center}
\begingroup
\scriptsize
\renewcommand{\arraystretch}{0.86}
\setlength{\tabcolsep}{2pt}
\setlength{\LTleft}{0pt plus 1fill}
\setlength{\LTright}{0pt plus 1fill}
\begin{longtable}[]{@{}llllll@{}}
\toprule
\textbf{Task formats} & \textbf{Abb.} & & \textbf{General level tasks} & \textbf{Specialized level tasks} & \textbf{Total} \\
\midrule
\endhead
True/False Questions & TFQ & Closed-form tasks & 493 & 199 & 692 \\
Single Choice Questions & SCQ & Closed-form tasks & 4265 & 270 & 4535 \\
Multiple Choice Questions & MCQ & Closed-form tasks & 70 & 52 & 122 \\
Objective Questions & OQ & Open-form tasks & 617 & 886 & 1503 \\
Reasoning Questions & RQ & Open-form tasks & 5 & 920 & 925 \\
\bottomrule
\end{longtable}
\endgroup
\begin{multicols}{2}

\subsection{Prompt for field labeling of ClimaBench}

The prompt for field labeling of ClimaBench was as follows:

\begingroup\scriptsize\setlength{\parindent}{0pt}
\par\smallskip\hrule\smallskip
System: You are an expert on climate policy.

Prompt:

I have a list that includes many types of subjects. Please determine which subject type the following question belongs to. You only need to answer with the subject type in the list.

If you think the question belongs to multiple subject types, please separate them with a ",".

Please answer the question in English.

If you think none of the above options apply, then answer with the closest subject type (it must be within the subject list I provided).

\textbf{\textless{} SUBJECT LIST \textgreater{}}

Question:\textbf{\textless QUESTION\textgreater{}}
\par\smallskip\hrule\smallskip\endgroup

Note: The \textbf{\textless{} SUBJECT LIST \textgreater{}} is provided in Supplement A2, and \textbf{\textless QUESTION\textgreater{}} refers to the question of tasks in ClimaBench, with corresponding labels available in Supplement D3.

\subsection{Disciplinary statistics of task formats in ClimaBench}

Table D2 shows the disciplinary frequency of five task formats tasks.

\end{multicols}
\begin{center}
\textbf{Table D2. Disciplinary frequency by task formats}
\end{center}
\begingroup
\scriptsize
\renewcommand{\arraystretch}{0.86}
\setlength{\tabcolsep}{2pt}
\setlength{\LTleft}{0pt plus 1fill}
\setlength{\LTright}{0pt plus 1fill}
\begin{longtable}[]{@{}llllllll@{}}
\toprule
\textbf{Subject area} & \textbf{Tier 1 Subject} & \textbf{Abb.} & \textbf{TFQ} & \textbf{SCQ} & \textbf{MCQ} & \textbf{OQ} & \textbf{RQ} \\
\midrule
\endhead
Physical Sciences & Chemical Engineering & ChE & 6 & 29 & 1 & 107 & 115 \\
& Chemistry & Chem & 31 & 247 & 0 & 26 & 29 \\
& Computer Science & CS & 3 & 13 & 0 & 13 & 1 \\
& Earth and Planetary Sciences & EPS & 198 & 1480 & 60 & 335 & 202 \\
& Energy & Energy & 102 & 429 & 31 & 137 & 5 \\
& Engineering & Eng & 131 & 233 & 8 & 62 & 52 \\
& Environmental Science & EnvSci & 250 & 2011 & 24 & 467 & 336 \\
& Materials Science & MatSci & 3 & 30 & 1 & 10 & 7 \\
& Mathematics & Math & 6 & 18 & 0 & 6 & 2 \\
& Physics and Astronomy & PhysAstro & 5 & 44 & 1 & 44 & 9 \\
Social Sciences & Arts and Humanities & ArtsHum & 0 & 3 & 3 & 69 & 7 \\
& Business, Management and Accounting & BMA & 20 & 54 & 0 & 79 & 78 \\
& Decision Sciences & DecSci & 7 & 17 & 0 & 16 & 17 \\
& Economics, Econometrics and Finance & Econ & 16 & 124 & 0 & 48 & 14 \\
& Psychology & Psych & 2 & 26 & 0 & 12 & 7 \\
& Social Sciences & SocSci & 111 & 801 & 0 & 55 & 21 \\
Life \& Health Sciences & Agricultural and Biological Sciences & AgBio & 91 & 1142 & 12 & 152 & 130 \\
& Medicine & Med & 46 & 334 & 0 & 6 & 1 \\
& Pharmacology, Toxicology and Pharmaceutics & PTP & 4 & 45 & 0 & 3 & 3 \\
\bottomrule
\end{longtable}
\endgroup
\begin{multicols}{2}

\section{Training Strategy for CFM}

\subsection{CPT strategy}

\textbf{Data balancing.} CPT was performed using the 78B-token interdisciplinary climate science corpus constructed in this study. To reduce potential distributional shift and maintain general reasoning ability, the training corpus was supplemented with non-climate foundational data. These additional data include mathematics, coding, and encyclopedic resources, which support logical structure, symbolic reasoning, and general expressive capacity. Specifically, the mathematics corpus incorporates OpenWebMath\numcite{57}, Algebraic Stack\numcite{58} and Math Instruct\numcite{59}; the code corpus included GitHub and Code Instruct datasets\numcite{60}; and the encyclopedia corpus included Fineweb Edu\numcite{61} and Wiki data\numcite{62}. The combined corpus was used during CPT to balance domain specialization with general-purpose reasoning competence.

\textbf{Training framework.} The training process was implemented using the open-source LlamaFactory framework\numcite{37}, in combination with DeepSpeed for distributed training\numcite{38}. To efficiently utilize computational resources, we adopted data parallelism and integrated the DeepSpeed ZeRO optimizer, which reduces memory consumption through parameter partitioning and offloading strategies. In addition, we employed Low-Rank Adaptation (LoRA)\numcite{63} to reduce parameter update overhead and accelerate training efficiency. Detailed descriptions of the pretraining settings and technique are provided in Supplement E3.

\textbf{Low-Rank Adaptation.} Low-Rank Adaptation (LoRA) is a parameter-efficient fine-tuning method that approximates weight updates using low-rank matrices parallel to the original model\numcite{63}. It is motivated by the observation that task-specific weight updates exhibit low-rank structure, allowing effective approximation via techniques such as singular value decomposition. By capturing most informative variations within a low-dimensional subspace, LoRA achieves strong fine-tuning performance with significantly reduced computational cost.

\subsection{SFT strategy}

\textbf{Data balancing.} We performed full-parameter instruction tuning on the CPT-adapted Qwen2.5-32B model using a mixture of domain-specific and general instructions. To mitigate catastrophic forgetting and preserve cross-domain reasoning ability, we adopted a mixed-data strategy, with 90\% of instructions from ClimaInstruct and 10\% from general-purpose datasets. Each sample consists of an instruction--response pair, and the training objective is to maximize the likelihood of generating accurate and contextually coherent responses.

\textbf{Training framework.} The SFT process was implemented using the LlamaFactory framework and incorporated several optimization techniques to improve training efficiency and stability, including Flash attention\numcite{39}, S\textsuperscript{2} attention\numcite{40}, and Mixed precision\numcite{41}. These techniques reduce memory consumption and accelerate training while maintaining numerical stability. Additional implementation details are reported in Supplement E3. To further stabilize training dynamics and improve generalization across downstream climate tasks, we applied model parameter averaging during fine-tuning.

\textbf{Model Parameter Averaging.} Parameter fusion is critical for optimization and deployment of large models, aiming to integrate multiple independently trained parameter modules or incremental updates into the base model. This process reduces computational overhead, memory usage, and model complexity during inference. Model Parameter Averaging\numcite{42}, a specific technique within parameter fusion, helps mitigate overfitting, enhance model stability, and improve out-of-domain performance. The core idea is to average or merge the parameters of multiple independently trained models according to their respective weights. This can be represented as:

\begin{equation}
W=\sum_{i=1}^{N}\alpha_i W_i
\tag{E1}
\end{equation}

where \(W_{i}\) denotes the parameters of the \(i\)-th model, and \(\alpha_{i}\) represents the weight assigned to the \(i\)-th model.

\subsection{Experimental environment and training settings}

Both CPT and SFT were conducted on a high-performance cluster comprising 4 NVIDIA HGX H800 nodes, each equipped with 8× H800-80G PCIe GPUs.

The CPT was conducted with the following hyperparameters: the initial training learning rate was set at 1e-5 and gradually reduced to 1e-6, the tensor parallel setting was 4, and the pipeline parallel setting was 8, with a context length limit of 1024 to ensure training stability.

The SFT was conducted with the following hyperparameters: learning rate was set at 1e-5, training for 1.0 epoch, a weight decay of 0.1, and a batch size of 2.

\subsection{Training validation}

\textbf{Validation on ClimaBench.} On ClimaBench, we evaluate the effects of CPT and SFT on climate-specific task performance across both general and specialized levels (Fig. E1). The combined model consistently outperforms the base model, with more pronounced improvements observed in specialized tasks, indicating enhanced capability in expert-level climate reasoning. Ablation analysis shows that CPT alone yields moderate gains, while SFT provides limited improvements when applied independently; the largest performance increase arises from their combination, suggesting complementary roles of domain adaptation and expert-aligned instruction tuning. These results validate that the training pipeline effectively improves climate-specific understanding.

\begin{minipage}{\linewidth}
\centering
\includegraphics[width=.88\linewidth,height=.28\textheight,keepaspectratio]{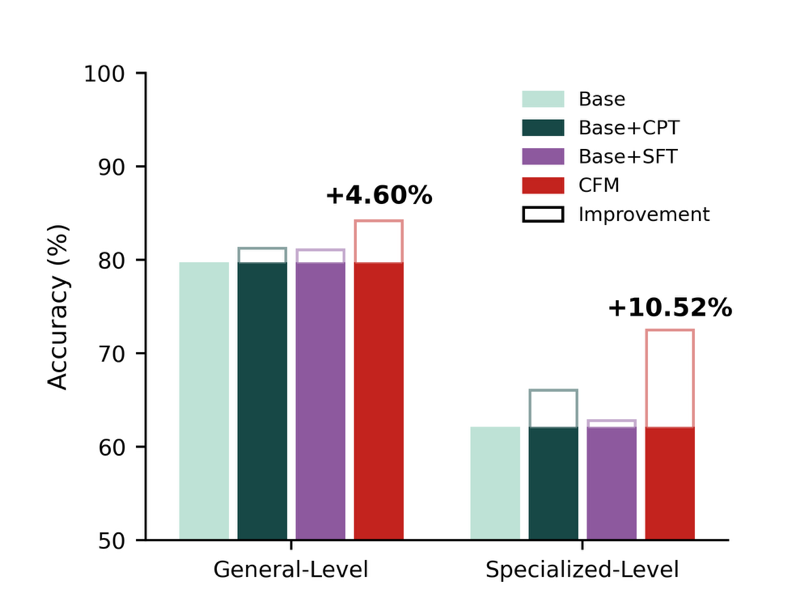}
\par\smallskip
\textbf{Figure E1. Performance gains from domain-specific training}
\end{minipage}

Note: Due to different evaluation methods, we present the weighted average accuracy of TFQ and SCQ.

\textbf{Validation on climate-related benchmarks.} To examine whether these improvements generalize beyond ClimaBench, we further evaluate the model on additional climate- and environment-related benchmarks, including CDP-QA and Pira 2.0 (Fig. E2). Compared with both general-purpose models and the domain-specific baseline ClimateGPT-70B, the model demonstrates consistent performance improvements across settings. This cross-benchmark evaluation supports that the gains reflect transferable climate knowledge rather than benchmark-specific adaptation.

\begin{minipage}{\linewidth}
\centering
\includegraphics[width=.88\linewidth,height=.28\textheight,keepaspectratio]{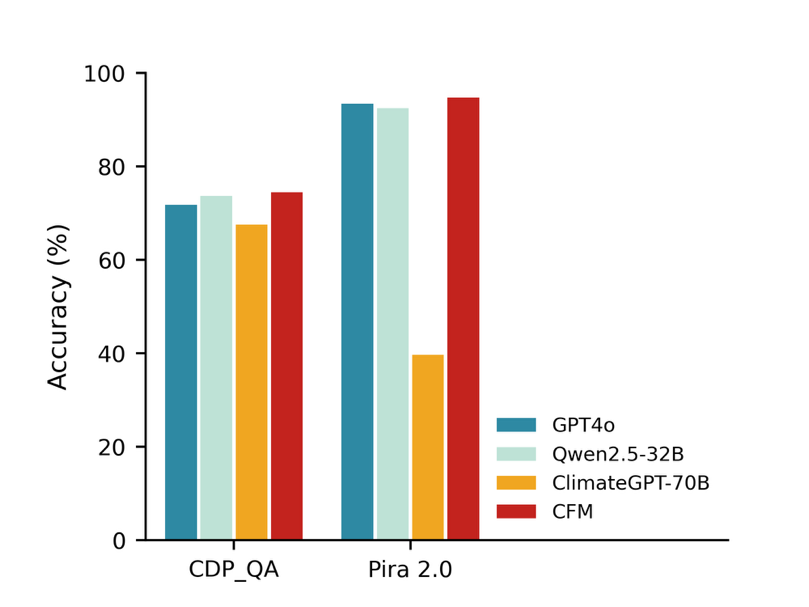}
\par\smallskip
\textbf{Figure E2. Performance on an external climate benchmark}
\end{minipage}

\textbf{Validation on general-domain benchmarks.} To assess whether domain adaptation affects general reasoning ability, we evaluate the model on CEval (Fig. E3). The results indicate that performance remains comparable to strong general-purpose models, while exceeding existing climate-specialized baselines. This suggests that the CPT and SFT strategy enhances domain-specific capabilities without compromising general reasoning.

\begin{minipage}{\linewidth}
\centering
\includegraphics[width=.88\linewidth,height=.28\textheight,keepaspectratio]{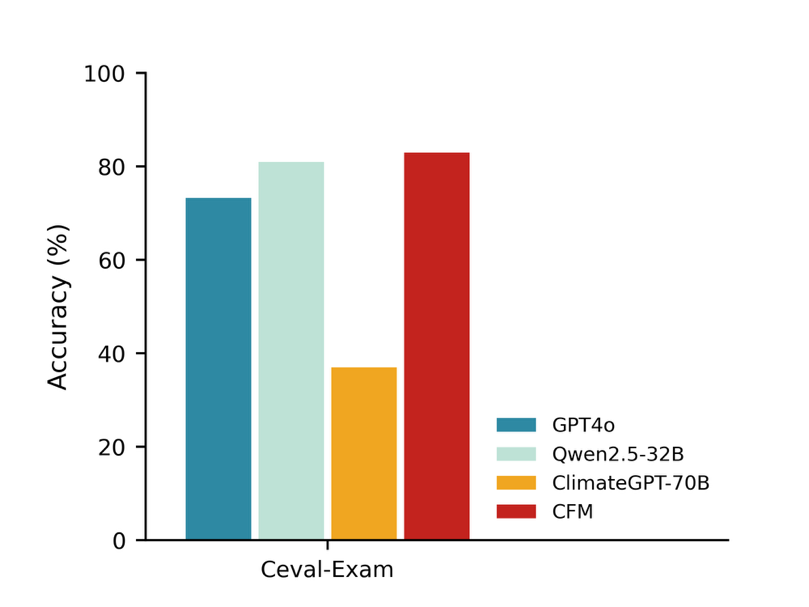}
\par\smallskip
\textbf{Figure E3. General benchmark performance}
\end{minipage}

\section{Interpreting Climate Knowledge through Embedding Analysis}

\subsection{Comparison of IPCC and general English word embeddings}

To examine how climate-specific training reshapes lexical representations, we analyze the embedding structure of climate-related and general English vocabularies. Figure F1a presents the embedding space of climate terms derived from the IPCC AR6 glossary. Compared to the base model, CFM exhibits more coherent clustering among climate-related words, indicating improved alignment within the climate-specific semantic subspace.

This reorganization is largely confined to climate-related vocabulary. As shown in Fig. F1b, the embedding distribution of the top 1,000 most frequent English words from the Google English word list remains stable after training, with no evidence of distortion in general lexical structure.

These results suggest that domain adaptation selectively refines representations within the climate subspace while preserving the overall semantic geometry of the model. This provides representation-level evidence that the observed performance gains arise from structured improvements in climate knowledge organization rather than global shifts in the embedding space.

\end{multicols}
\begin{center}
\begin{minipage}{.95\textwidth}
\centering
\includegraphics[width=.88\textwidth,height=.50\textheight,keepaspectratio]{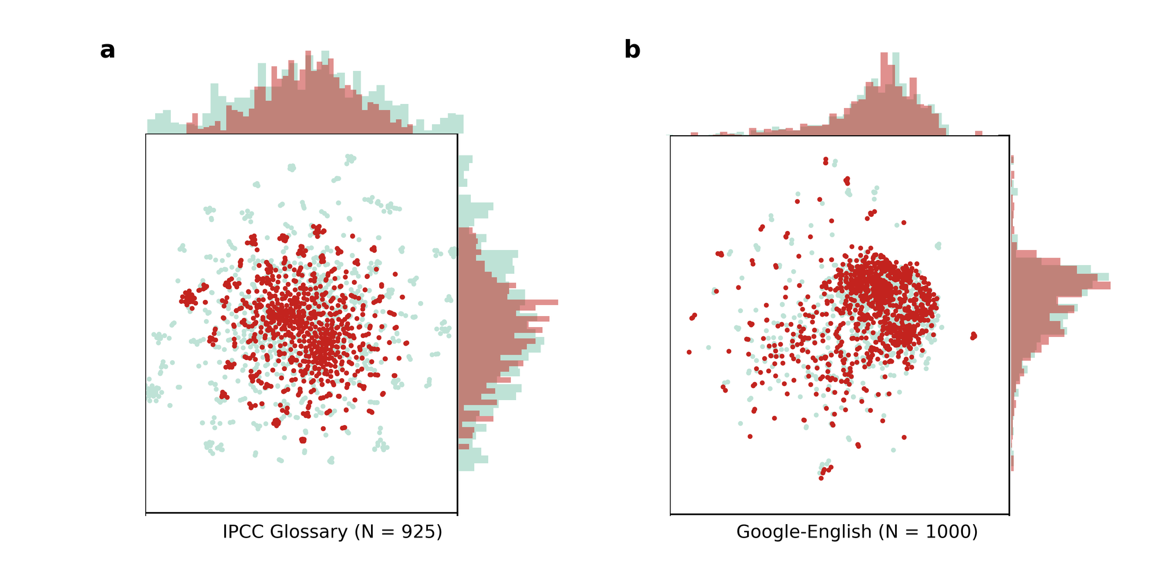}
\par\smallskip
\textbf{Figure F1. Comparison of IPCC and general English word embeddings}
\end{minipage}
\end{center}
\begin{multicols}{2}

\subsection{Changes in semantic topology of word embeddings}

To further characterize how climate-specific training reshapes lexical structure, we analyze pairwise distances between word embeddings before and after training. A uniform contraction or expansion of distances would indicate a lack of structural learning. Instead, we observe heterogeneous shifts that reflect meaningful reorganization in the embedding space. Representative examples illustrate these changes (Fig. F2). The distance between warming and ocean decreases, consistent with their strong coupling in climate processes such as ocean heat uptake and circulation feedbacks. Similarly, energy and climate become closer, reflecting improved alignment with core concepts such as energy balance and radiative forcing. In contrast, the distance between sea and ocean increases, indicating enhanced semantic differentiation between related but distinct concepts.

Overall, these patterns suggest that climate-specific training induces systematic changes in semantic topology, bringing cross-disciplinary but mechanistically related concepts closer while separating concepts with distinct meanings.

\end{multicols}
\begin{center}
\begin{minipage}{.95\textwidth}
\centering
\includegraphics[width=.88\textwidth,height=.50\textheight,keepaspectratio]{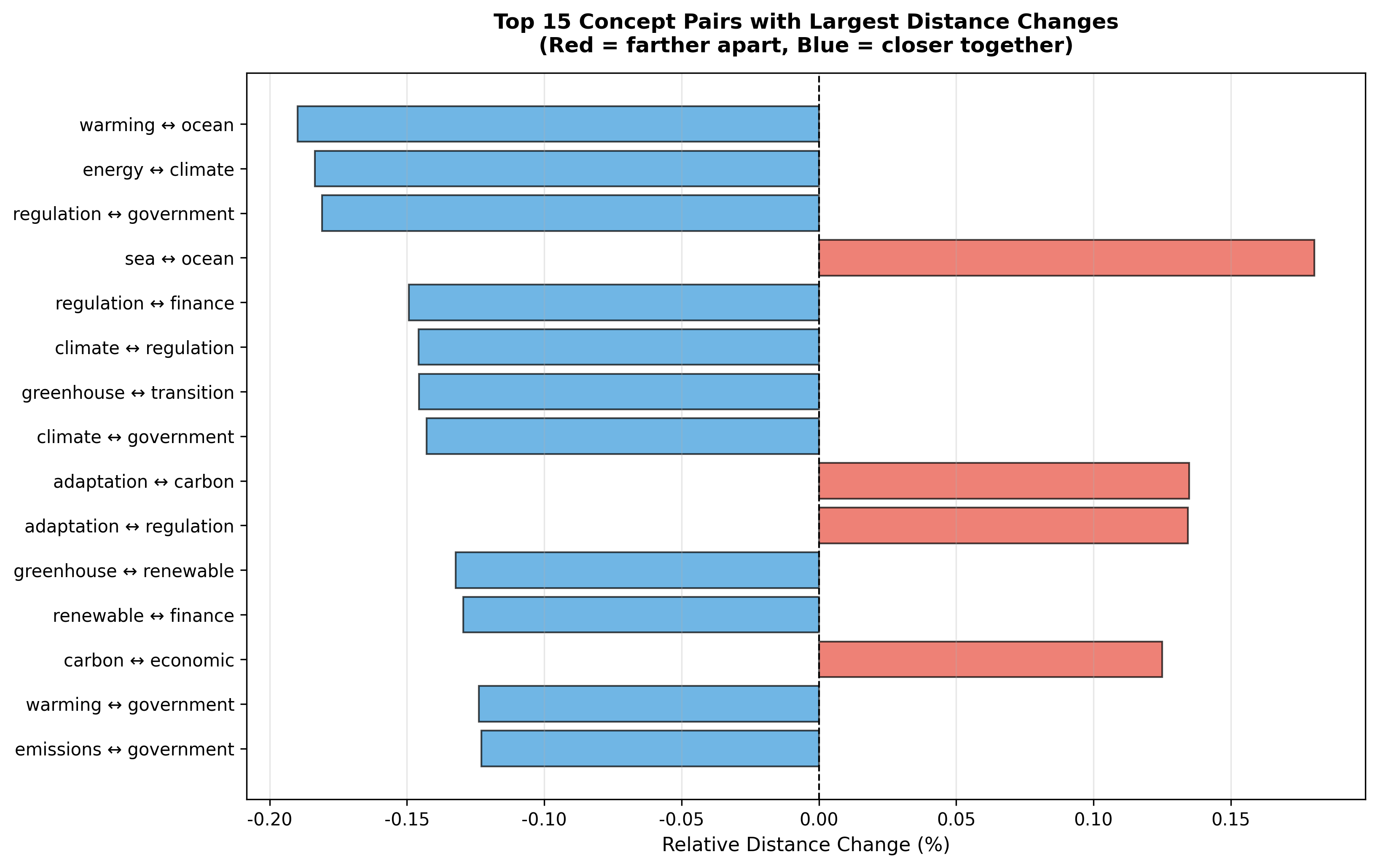}
\par\smallskip
\textbf{Figure F2. Changes in semantic topology of word embeddings}
\end{minipage}
\end{center}
\begin{multicols}{2}

\subsection{Consistency of relational chains in embedding space}

To examine whether the model captures structured relationships across multiple steps, we analyze semantic ``relational chains'' that link physical processes, intermediate events, and downstream impacts. In climate systems, such chains often take the form of multi-stage dependencies (e.g., ocean heat content increase → rapid cyclone intensification → port disruption losses), spanning physical, hazard, and socioeconomic domains.

To quantify this property, we introduce a consistency metric based on distances in the embedding space. For a relational chain \(A \rightarrow B \rightarrow C\), we define a monotonic gradient condition requiring that the intermediate step is closer to both endpoints than the endpoints are to each other, i.e., \(d(A,B) < d(A,C)\)and \(d(B,C) < d(A,C)\). This condition ensures that the chain is represented as a directed and coherent progression in the latent space.

We construct representative relational chains across multiple climate domains, including Physical → Impact, Socioeconomic → Forcing, Scenario → Transition, and Risk → Behavior categories (Table F1). The results show that CFM satisfies the monotonic gradient condition more frequently than baseline models, particularly in Socioeconomic → Forcing and Scenario → Transition chains. This indicates improved preservation of long-range relational structure, suggesting that the model better integrates physical processes, economic dynamics, and policy outcomes within a unified semantic space.

\end{multicols}
\begin{center}
\textbf{Table F1. Consistency of relational chains in embedding space}
\end{center}
\begingroup
\scriptsize
\renewcommand{\arraystretch}{0.86}
\setlength{\tabcolsep}{2pt}
\setlength{\LTleft}{0pt plus 1fill}
\setlength{\LTright}{0pt plus 1fill}
\begin{longtable}[]{@{}llll@{}}
\toprule
\endhead
\textbf{Field} & \textbf{Qwen2.5-32B} \textbf{(Base)} & \textbf{Fuxi-CFM} & \textbf{Improvement} \\
Physical → Impact & 41.12\% & 42.50\% & 3.16\% \\
Socioeconomic → Forcing & 22.19\% & 25.00\% & 12.22\% \\
Scenario → Transition & 42.62\% & 47.00\% & 10.27\% \\
Risk → Behavior & 43.69\% & 47.50\% & 8.72\% \\
\bottomrule
\end{longtable}
\endgroup
\begin{multicols}{2}

Note: Each category comprises 200 manually annotated relational chains. Results are computed based on 100 Monte Carlo random trials.

\section{Comprehensive Interdisciplinary Evaluation}

\subsection{Performance across models by task type}

Figure. G1 compares model performance across task types, including TFQ, SCQ, MCQ, OQ, and RQ, for both general-level and specialized-level settings. The evaluation includes general-purpose LLMs, climate-specialized models, and subdomain-focused climate models (e.g., OceanGPT and MiningGPT), providing a broad baseline for comparison.

Panels a--d report performance on TFQ, SCQ, MCQ, and OQ, respectively, where models are ranked according to specialized-level results, with general-level performance shown alongside. Panel \textbf{e} presents reasoning task performance, where models are ranked based on reasoning outcome scores, with reasoning process consistency reported separately. Across task types, this figure provides a unified comparison of model behavior under both general and specialized settings, highlighting differences in performance across formats and model categories.

\end{multicols}
\begin{center}
\begin{minipage}{.95\textwidth}
\centering
\includegraphics[width=.88\textwidth,height=.50\textheight,keepaspectratio]{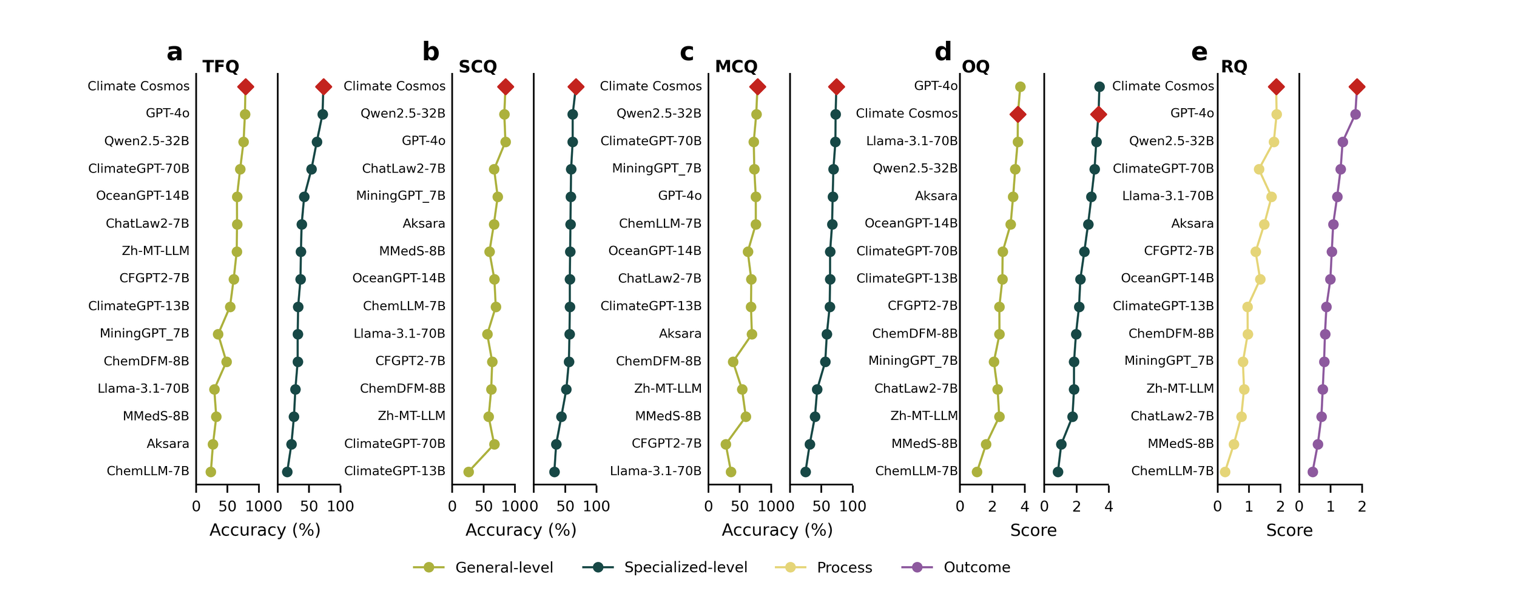}
\par\smallskip
\textbf{Figure G1. Performance comparison across models and task types}
\end{minipage}
\end{center}
\begin{multicols}{2}

\subsection{CFM interdisciplinary stress test by task type}

This analysis is conducted separately for specialized-level and general-level tasks, and further stratified by task type, including true--false (TFQ), single-choice (SCQ), multiple-choice (MCQ), open-ended questions (OQ), and reasoning questions (RQ).

Figure. G2 presents model performance as a function of disciplinary breadth (x-axis), with the y-axis reporting accuracy for objective tasks (TFQ, SCQ, MCQ) and scores for open-ended (OQ) and reasoning tasks. Results are shown separately for specialized-level and general-level tasks across different task formats. Panels a--d show specialized-level performance for TFQ, SCQ, MCQ, and OQ, respectively, across increasing disciplinary breadth. Panel e reports reasoning process consistency under the same setting. Panels f--i present general-level performance for TFQ, SCQ, MCQ, and OQ, respectively, and panel j reports reasoning outcome consistency.

Across all settings, performance is evaluated consistently along the same horizontal axis of disciplinary breadth, enabling direct comparison of how model behavior changes as tasks require broader cross-domain integration.

\end{multicols}
\begin{center}
\begin{minipage}{.95\textwidth}
\centering
\includegraphics[width=.88\textwidth,height=.50\textheight,keepaspectratio]{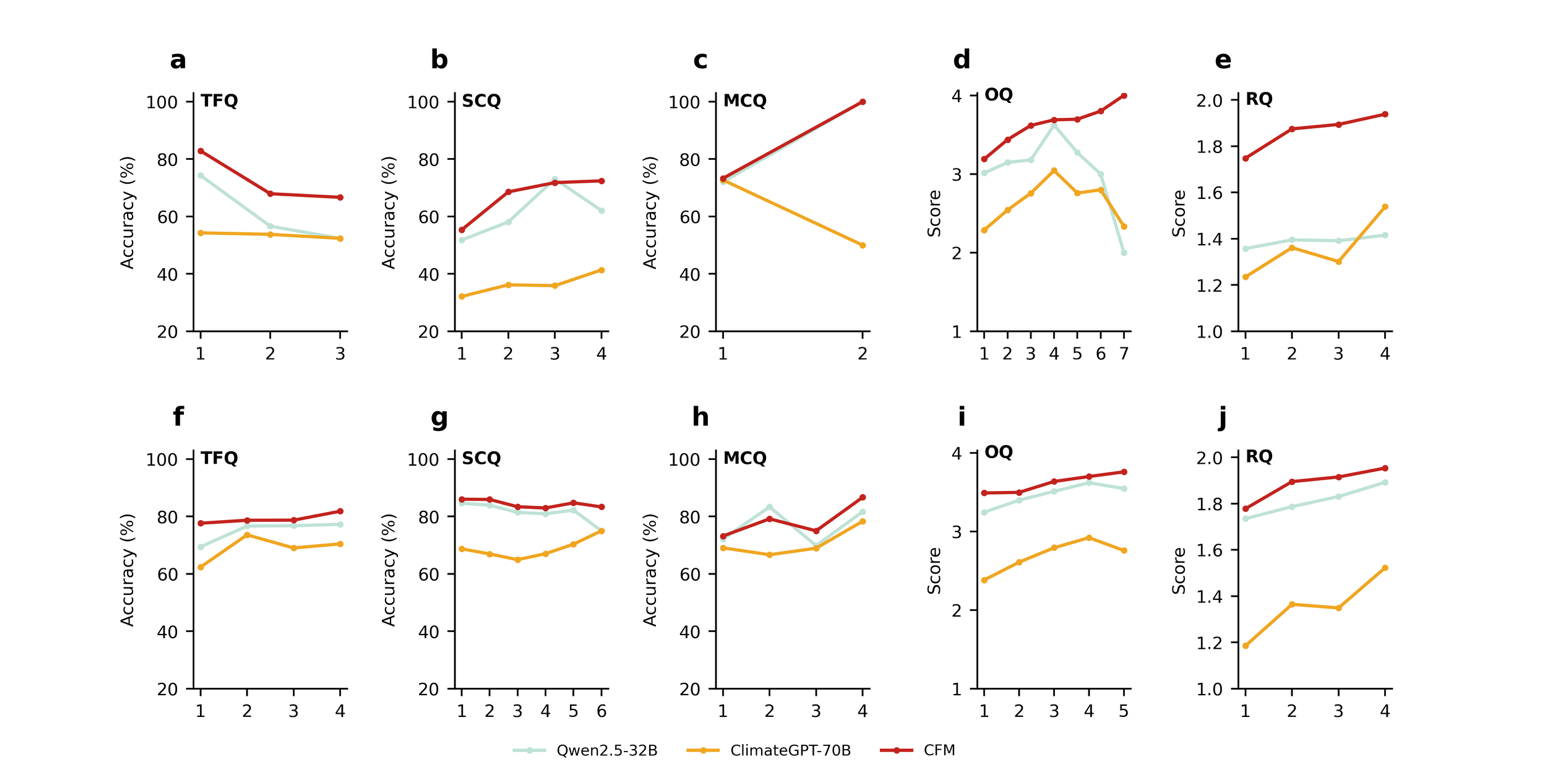}
\par\smallskip
\textbf{Figure G2. Performance across disciplinary breadth by task type}
\end{minipage}
\end{center}
\begin{multicols}{2}

\end{multicols}
\clearpage
\subsection{Disciplinary distribution of LLMs for comparison}
\begin{center}
\textbf{Table G1. Disciplinary distribution of LLMs}\par\medskip
\end{center}
\begingroup\scriptsize\setlength{\tabcolsep}{2pt}
\begin{center}\textbf{Panel 1: ClimateGPT-13B--Qwen-32B}\end{center}
\begin{tabularx}{\textwidth}{@{}p{.14\textwidth}X*{5}{>{\centering\arraybackslash}p{.085\textwidth}}@{}}
\toprule
\textbf{Field} & \textbf{Discipline} & \shortstack{\bfseries Climate\\\bfseries GPT 13B} & \shortstack{\bfseries Climate\\\bfseries GPT 70B} & \shortstack{\bfseries Llama3.1} & \shortstack{\bfseries Qwen\\72B} & \shortstack{\bfseries Qwen\\32B} \\
\midrule
Physical Sciences & Chemical Engineering & √ & √ & √ & √ & √ \\
 & Chemistry & √ & √ & √ & √ & √ \\
 & Computer Science & √ & √ & √ & √ & √ \\
 & Earth and Planetary Sciences & √ & √ & √ & √ & √ \\
 & Energy & √ & √ & √ & √ & √ \\
 & Engineering & √ & √ & √ & √ & √ \\
 & Environmental Science & √ & √ & √ & √ & √ \\
 & Materials Science & √ & √ & √ & √ & √ \\
 & Mathematics & √ & √ & √ & √ & √ \\
 & Physics and Astronomy & √ & √ & √ & √ & √ \\
Social Sciences & Arts and Humanities & √ & √ & √ & √ & √ \\
 & Business, Management and Accounting & √ & √ & √ & √ & √ \\
 & Decision Sciences & √ & √ & √ & √ & √ \\
 & Economics, Econometrics and Finance & √ & √ & √ & √ & √ \\
 & Psychology & √ & √ & √ & √ & √ \\
 & Social Sciences & √ & √ & √ & √ & √ \\
Life \& Health Sciences & Agricultural and Biological Sciences & √ & √ & √ & √ & √ \\
 & Medicine & √ & √ & √ & √ & √ \\
 & Pharmacology, Toxicology and Pharmaceutics & √ & √ & √ & √ & √ \\
\bottomrule
\end{tabularx}
\medskip
\begin{center}\textbf{Panel 2: GPT-4o--ChemLLM}\end{center}
\begin{tabularx}{\textwidth}{@{}p{.14\textwidth}X*{5}{>{\centering\arraybackslash}p{.085\textwidth}}@{}}
\toprule
\textbf{Field} & \textbf{Discipline} & \shortstack{\bfseries GPT-4o} & \shortstack{\bfseries CFGPT2} & \shortstack{\bfseries ChatLaw2} & \shortstack{\bfseries ChemDFM} & \shortstack{\bfseries ChemLLM} \\
\midrule
Physical Sciences & Chemical Engineering & √ &  &  & √ & √ \\
 & Chemistry & √ &  &  & √ & √ \\
 & Computer Science & √ &  &  &  &  \\
 & Earth and Planetary Sciences & √ &  &  &  &  \\
 & Energy & √ &  &  & √ & √ \\
 & Engineering & √ &  &  & √ & √ \\
 & Environmental Science & √ &  &  & √ & √ \\
 & Materials Science & √ &  &  & √ & √ \\
 & Mathematics & √ &  &  &  &  \\
 & Physics and Astronomy & √ &  &  &  &  \\
Social Sciences & Arts and Humanities & √ & √ &  &  &  \\
 & Business, Management and Accounting & √ &  & √ &  &  \\
 & Decision Sciences & √ & √ &  &  &  \\
 & Economics, Econometrics and Finance & √ & √ &  &  &  \\
 & Psychology & √ &  &  &  &  \\
 & Social Sciences & √ & √ &  &  &  \\
Life \& Health Sciences & Agricultural and Biological Sciences & √ &  &  &  &  \\
 & Medicine & √ &  &  &  &  \\
 & Pharmacology, Toxicology and Pharmaceutics & √ &  &  &  &  \\
\bottomrule
\end{tabularx}
\medskip
\begin{center}\textbf{Panel 3: Cropin Aksara--Zh-MT-LLM}\end{center}
\begin{tabularx}{\textwidth}{@{}p{.14\textwidth}X*{5}{>{\centering\arraybackslash}p{.085\textwidth}}@{}}
\toprule
\textbf{Field} & \textbf{Discipline} & \shortstack{\bfseries Cropin\\Aksara} & \shortstack{\bfseries MiningGPT} & \shortstack{\bfseries MMedS} & \shortstack{\bfseries OceanGPT} & \shortstack{\bfseries Zh-MT-\\LLM} \\
\midrule
Physical Sciences & Chemical Engineering &  &  & √ &  &  \\
 & Chemistry &  &  & √ &  &  \\
 & Computer Science &  &  &  &  &  \\
 & Earth and Planetary Sciences & √ &  &  & √ &  \\
 & Energy &  & √ &  & √ &  \\
 & Engineering &  &  &  & √ &  \\
 & Environmental Science & √ & √ &  & √ &  \\
 & Materials Science &  &  & √ &  &  \\
 & Mathematics &  &  &  &  &  \\
 & Physics and Astronomy & √ &  &  & √ & √ \\
Social Sciences & Arts and Humanities &  &  &  &  &  \\
 & Business, Management and Accounting &  &  &  &  & √ \\
 & Decision Sciences &  &  &  &  &  \\
 & Economics, Econometrics and Finance &  &  &  &  &  \\
 & Psychology &  &  & √ &  &  \\
 & Social Sciences &  &  &  &  &  \\
Life \& Health Sciences & Agricultural and Biological Sciences & √ &  &  &  &  \\
 & Medicine &  &  & √ &  &  \\
 & Pharmacology, Toxicology and Pharmaceutics &  &  & √ &  &  \\
\bottomrule
\end{tabularx}
\endgroup
\begin{multicols}{2}

\subsection{Criteria of open-ended questions evaluation}

We adopted a 4-scale Likert scoring scheme to assess response quality of open-ended questions. The following are specific scoring criteria:

"0": "ERROR - The generated text deviates significantly from the Reference Answer."

"1": "Somewhat helpful - Compared with the Reference Answer, the Response Answer has some relevance to the user's question, but it may be unclear or incomplete. It provides only partial information, or the information provided may not be useful for the user's needs."

"2": "Moderately helpful - The response is relevant to the user's question, and it provides a clear and complete answer. However, compared with the Reference Answer, it may lack detail or explanation that would be helpful for the user."

"3": "Helpful - The main content of Reference Answer and Response is basically the same."

"4": "Highly helpful - Compared with the Reference Answer, The Response Answer provides a clear, complete, and detailed answer. It offers additional information or explanations that are not only useful but also insightful and valuable to the user. "

\subsection{Standard of reasoning questions evaluation}

Both the reasoning process and reasoning outcomes are scored on a three-point scale, inspired by the recent study published in the Nature\numcite{47}. The following are specific scoring criteria:

'0': Compared to the reference answer, the Response Answer significantly deviates in meaning, contains obvious errors, or includes content that is inconsistent with human values.

'1': Compared to the reference answer, the Response Answer is somewhat relevant to the question but may be unclear or incomplete. It only provides partial information, or the information provided may not be helpful to the user's needs.

'2': The main content of the Response is essentially the same as the reference answer. Alternatively, compared to the reference answer, the Response provides a clear, complete, and detailed answer.

\section{Comprehensive Evaluation on the 220 Tasks}

\subsection{Reasoning analytical components across climate domains}

\end{multicols}
\begin{center}
\textbf{Table H1. Expected analytical components}
\end{center}
\begingroup
\scriptsize
\renewcommand{\arraystretch}{0.86}
\setlength{\tabcolsep}{2pt}
\setlength{\LTleft}{0pt plus 1fill}
\setlength{\LTright}{0pt plus 1fill}
\begin{longtable}[]{@{}llll@{}}
\toprule
\textbf{Class} & \textbf{Thematic Focus} & \textbf{Expected Analytical Components} & \textbf{Phase} \\
\midrule
\endhead
Class 1 & Mitigation \& Energy Transitions & Emissions profile and constraint context & I \\
& & Mitigation levers and transition pathways & II \\
& & System interactions and dependencies & II \\
& & Trade-offs and co-benefits & II \\
& & Pathway credibility and transition risks & III \\
Class 2 & Adaptation \& Climate Risk Management & Risk framing and exposure context & I \\
& & Vulnerability and sensitivity drivers & II \\
& & Adaptation strategy space & II \\
& & Trade-offs, maladaptation and lock-in risks & II \\
& & Implementation, monitoring and learning & III \\
Class 3 & Climate--Health--Food Systems & Climate stressors and affected systems & I \\
& & Impact pathways and biological/system mechanisms & II \\
& & Cross-system interactions & II \\
& & Distributional and compound effects & II \\
& & Integrated response implications & III \\
Class 4 & Carbon Removal \& Accounting & Role in mitigation architecture & I \\
& & Measurement and accounting basis & II \\
& & Equivalence and substitution logic & II \\
& & Environmental and social side effects & II \\
& & Governance and integrity safeguards & III \\
Class 5 & Governance, Finance \& Political Economy & Governance arena and actor landscape & I \\
& & Incentives and political economy drivers & II \\
& & Distributional consequences & II \\
& & Systemic financial and macroeconomic risks & II \\
& & Institutional design and reform pathways & III \\
Class 6 & Climate System Dynamics and Earth Processes & Baseline state and observed change & I \\
& & Forcing agents and internal variability & II \\
& & Process linkages and feedback structure & II \\
& & Nonlinearities and tipping behavior & II \\
& & Implications for future trajectories & III \\
Class 7 & Equity, Justice \& Societal Transformation & Normative principles and justice framework & I \\
& & Existing inequality structures & II \\
& & Distributional impacts of climate action/inaction & II \\
& & Trade-offs between efficiency and justice & II \\
& & Transformative pathways and social legitimacy & III \\
\bottomrule
\end{longtable}
\endgroup
\begin{multicols}{2}

\subsection{Measurement of reasoning consistency across phases}

\textbf{Overview.} We introduce a breakpoint-based diagnostic method to quantify reasoning consistency across phases. The method identifies where a model's analytical reasoning chain fails when addressing multi-step climate questions. It is implemented as a second-stage analysis built upon component coverage evaluation: an evaluator model first determines which predefined analytical components are semantically addressed in a response, after which deterministic rules extract structural features indicating breakdown points. This two-stage design avoids inconsistencies associated with repeated LLM inference within a single evaluation pipeline.

\textbf{Theoretical Background.} In climate science, high-quality responses to open-ended questions typically require addressing multiple interrelated analytical dimensions. These analytical components exhibit a natural logical progression: earlier components (e.g., emission scenario analysis) provide the reasoning foundation for subsequent ones (e.g., mitigation pathways).

Inspired by fine-grained atomic evaluation methods such as FActScore\numcite{64}, we decompose the expected model response into an ordered sequence of analytical components. However, unlike factual verification, our method focuses on the structural completeness of the analytical reasoning chain rather than the correctness of individual facts. This approach also draws on Integrative Complexity Theory\numcite{65}, which provides a framework for assessing multi-dimensional analysis and inter-dimensional integration capabilities in text.

\textbf{Method Design.} The method employs a two-stage evaluation architecture:

i) Stage 1: Component Coverage Detection (LLM-based). An evaluator model (e.g., GPT series) performs semantic-level coverage assessment, determining which of the predefined analytical components are addressed in the model's response (see more details in Supplement H6). Semantic matching is adopted rather than exact keyword matching\numcite{66}.

ii) Stage 2: Breakpoint Feature Extraction (Rule-based). Based on the coverage results from Stage 1, deterministic rules compute breakpoint metrics. The decision to decouple these two stages is motivated by: (1) avoiding inconsistency from dual LLM inference within the same evaluation pipeline\numcite{67}; (2) ensuring reproducibility and interpretability of breakpoint metrics; and (3) reducing evaluation costs.

\textbf{Formal definition.} Given a question \(q\) with an ordered component set \(C = \{ c_{1},c_{2},...,c_{k}\}\), and a model response \(a_{m,q}\), the coverage detection yields a subset \(C^{+} \subseteq C\). The breakpoint set is defined as the uncovered components:

\begin{equation}
B=C\setminus C^{+}=\{c_i\mid c_i\notin C^{+}\}
\tag{H1}\label{eq:H1}
\end{equation}

From this, we extract the following diagnostic metrics as shown in Table H2:

\end{multicols}
\begin{center}
\textbf{Table H2. Diagnostic metrics}
\end{center}
\begingroup
\scriptsize
\renewcommand{\arraystretch}{0.86}
\setlength{\tabcolsep}{2pt}
\setlength{\LTleft}{0pt plus 1fill}
\setlength{\LTright}{0pt plus 1fill}
\begin{longtable}[]{@{}
  >{\raggedright\arraybackslash}p{(\columnwidth - 4\tabcolsep) * \real{0.33}}
  >{\raggedright\arraybackslash}p{(\columnwidth - 4\tabcolsep) * \real{0.33}}
  >{\raggedright\arraybackslash}p{(\columnwidth - 4\tabcolsep) * \real{0.33}}@{}}
\toprule
\textbf{Metric} & \textbf{Definition} & \textbf{Description} \\
\midrule
\endhead
Coverage Rate \(R_{\text{cov}}\) & \(R_{\text{cov}} = \frac{C^{+}}{C}\) & Proportion of components covered \\
Breakpoint Count \(N_{b}\) & \(N_{b} = \left| B \right|\) & Total number of uncovered components \\
First Breakpoint Position \(P_{\text{first}}\) & \(P_{\text{first}} = \min_{c_{i} \in B}c_{i}\)

(None if \(B = \varnothing\)) & Position of the first uncovered component in the ordered sequence \\
Breakpoint Position Set \(P_{b}\) & \(P_{b} = i\ \left| c_{i} \in B \right.\ \) & Set of all breakpoint positions \\
\bottomrule
\end{longtable}
\endgroup
\begin{multicols}{2}

We define the primary diagnostic metric \(P_{\text{first}}\) as the index of the first missing component in the sequence. This metric characterizes the depth at which the reasoning chain first breaks down: lower values indicate early-stage failure, whereas higher values indicate that the model sustains reasoning through earlier phases and fails only at deeper stages.

\subsection{Discipline-wise consistency across reasoning phases}

Figure H1 shows the discipline-wise stability of climate reasoning with increasing analytical depth.

\end{multicols}
\begin{center}
\begin{minipage}{.95\textwidth}
\centering
\includegraphics[width=.88\textwidth,height=.50\textheight,keepaspectratio]{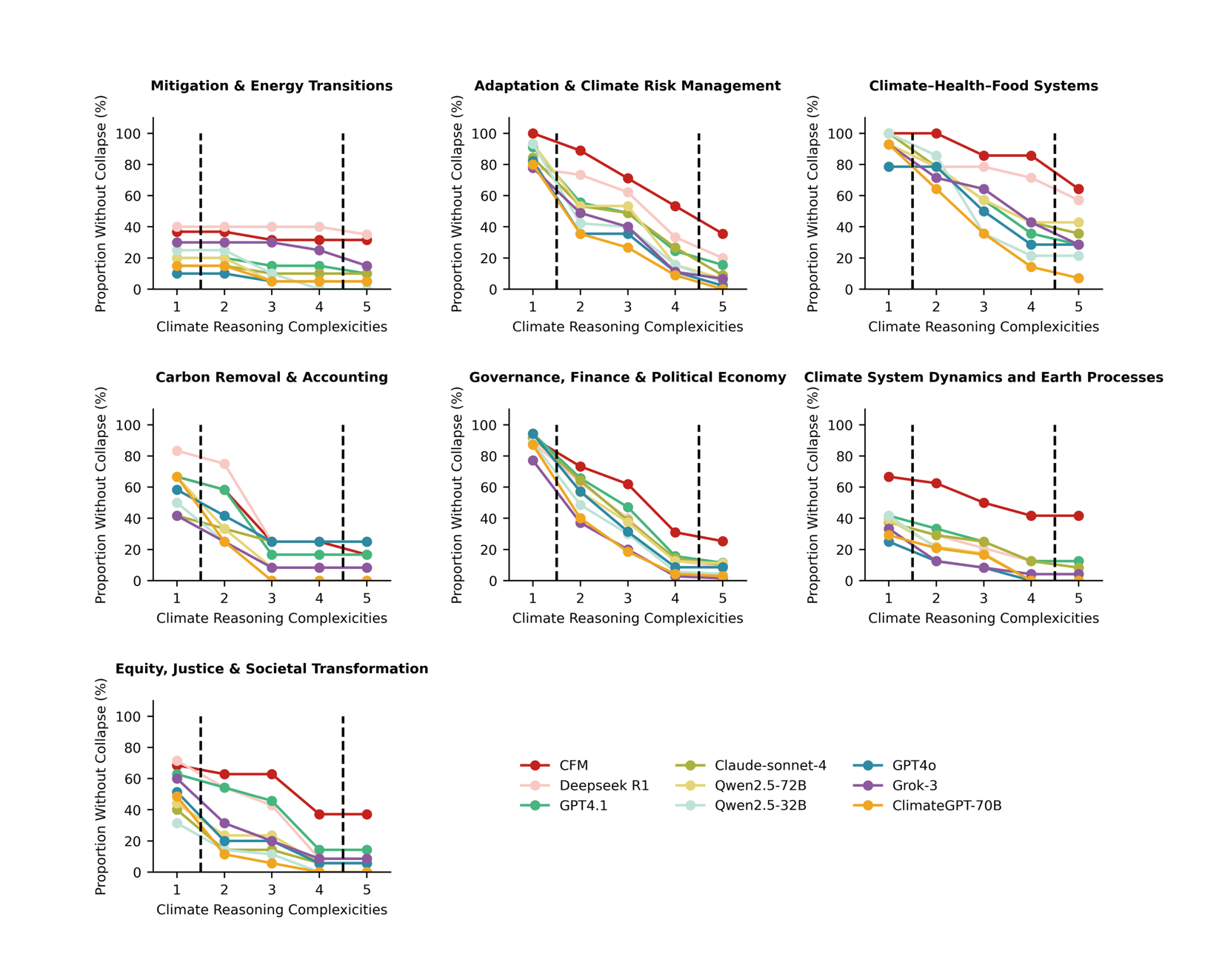}
\par\smallskip
\textbf{Figure H1. Discipline-wise stability of climate reasoning with increasing analytical depth}
\end{minipage}
\end{center}
\begin{multicols}{2}

\subsection{IPCC semantic similarity computation}

We quantify semantic alignment between model-generated responses and IPCC assessment reports using an embedding-based similarity metric.

All IPCC reports are concatenated and segmented into 94,928 text chunks (approximately 2,000 characters each). Each chunk is encoded into a semantic vector using the \emph{text-embedding-3-large} model\numcite{68}, forming a vector database that represents the IPCC knowledge space.

For each model response, we compute its embedding using the same model and retrieve the Top-\emph{k} most similar IPCC text chunks based on cosine similarity. The semantic similarity score is defined as the average cosine similarity between the response embedding and the retrieved chunks, evaluated at multiple retrieval depths (\emph{k} = 3, 5, 10, 15, 20).

\subsection{Coverage of domain-specific terminology}

To assess the alignment of model outputs with established climate science terminology, we conduct a lexical analysis based on the IPCC AR6 glossary. For each task, we count the number of glossary terms appearing in model responses and compare this coverage with that of general-purpose and climate-oriented baselines.

Figure H2 shows that CFM consistently incorporates a higher number of IPCC-aligned terms per task. This indicates closer alignment with the standardized conceptual vocabulary used in authoritative climate assessments, reflecting more precise and domain-consistent language use.

These findings suggest that climate-specific training improves not only task performance but also the model's ability to anchor its responses in established scientific terminology, providing additional evidence of enhanced domain expertise.

\begin{minipage}{\linewidth}
\centering
\includegraphics[width=.95\linewidth,height=.32\textheight,keepaspectratio]{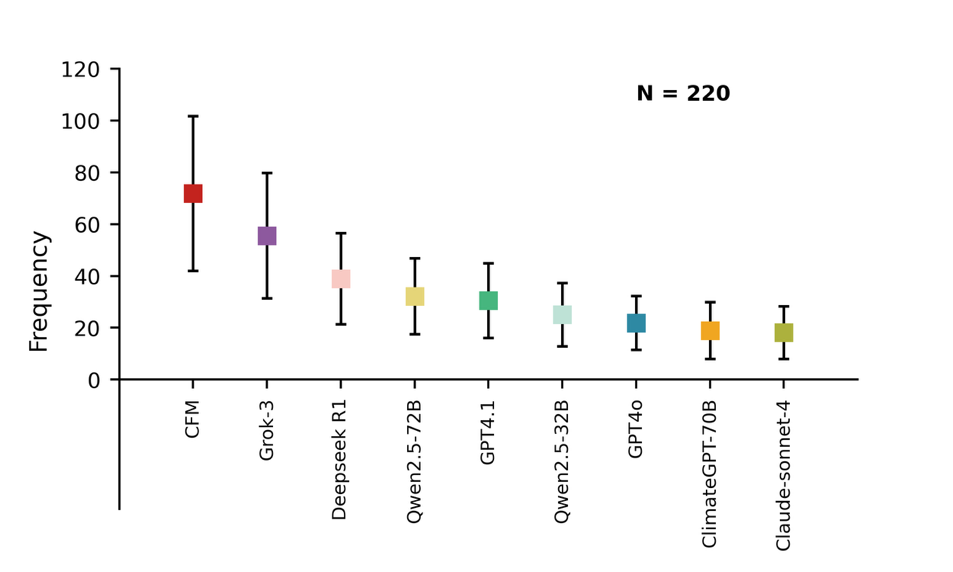}
\par\smallskip
\textbf{Figure H2. Coverage of domain-specific terminology}
\end{minipage}

\subsection{Reasoning steps coverage evaluation prompt}

The prompt for evaluating coverage of reasoning steps in models' responses was as follows:

\begingroup\scriptsize\setlength{\parindent}{0pt}
\par\smallskip\hrule\smallskip
You are an expert evaluator assessing the coverage of Expected Analytical Components in a climate model's response.

**Question Class \{class\_num\}: \{thematic\_focus\}**

**Question:**

\{question\}

**Expected Analytical Components (5 total):**

\{components\_list\}

**Model's Answer:**

\{answer\}

**Evaluation Task:**

Determine which of the 5 Expected Analytical Components above are covered in the model's answer.

"Covered" means the answer addresses the component semantically (not requiring exact keyword match). The components should be evaluated in order.

**Output Format:**

Provide ONLY a JSON object with this exact format:

\{

"covered\_components": {[}1, 2, 3{]}, // List of component numbers (1-5) that are covered. Use empty list {[}{]} if none covered.

"reasoning": "Brief explanation of which components are covered and why (2-3 sentences)"

\}

Example 1: If components 1, 2, and 4 are covered but 3 and 5 are not:

\{

"covered\_components": {[}1, 2, 4{]},

"reasoning": "The answer addresses emission constraints and mitigation pathways but misses system interactions and transition risks."

\}

Example 2: If only component 1 is covered:

\{

"covered\_components": {[}1{]},

"reasoning": "The answer only discusses the emission profile context but lacks analysis of other components."

\}
\par\smallskip\hrule\smallskip\endgroup

\subsection{Trade-off evaluation prompt}

The prompt for evaluating trade-off analysis in models' responses was as follows:

\begingroup\scriptsize\setlength{\parindent}{0pt}
\par\smallskip\hrule\smallskip
You are an expert evaluator assessing the quality of climate policy and science responses.

Your task is to evaluate how well the model's answer addresses the SPECIFIC trade-off that is central to this question.

**Question:**

\{question\}

**EXPECTED Trade-off (defined in the question classification):**

\{key\_tradeoff\}

**Model's Answer:**

\{answer\}

**Evaluation Instructions:**

Focus specifically on whether the model's answer addresses the EXPECTED trade-off listed above.

**Evaluation Criteria for Trade-off Articulation:**

Score 0 - No Discussion of Expected Trade-off:

- The answer does NOT mention the expected trade-off ("\{key\_tradeoff\}")

- The response treats the problem as having a straightforward solution without considering this tension

- OR discusses other trade-offs but completely misses the expected one

Score 1 - Mentions Expected Trade-off but Lacks Analysis:

- The answer acknowledges the expected trade-off exists (mentions "\{key\_tradeoff\}" or its components)

- However, it does NOT analyze:

* How these competing objectives interact or conflict

* The mechanisms through which the trade-off operates

* The implications/consequences of choosing one side over the other

- Merely naming the tension without substantive analysis

Score 2 - Thorough Analysis of Expected Trade-off:

- The answer explicitly identifies and discusses the expected trade-off ("\{key\_tradeoff\}")

- Provides substantive analysis of how these competing objectives interact

- Explains the consequences, tensions, or impacts involved

- May offer strategies for balancing, navigating, or mitigating the trade-off

**Output Format:**

Provide ONLY a JSON object in the following format (no other text):

\{

"score": 0 or 1 or 2,

"reasoning": "Brief explanation focusing on whether and how well the expected trade-off ('\{key\_tradeoff\}') was addressed"

\}
\par\smallskip\hrule\smallskip\endgroup

\subsection{Uncertainty handling evaluation prompt}

The prompt for uncertainty handling in models' responses was as follows:

\begingroup\scriptsize\setlength{\parindent}{0pt}
\par\smallskip\hrule\smallskip
You are an expert evaluator assessing how climate models handle uncertainty in their responses.

Your task is to evaluate whether the model's answer acknowledges uncertainty and incorporates it into the analysis.

**Question:**

\{question\}

**Model's Answer:**

\{answer\}

**Evaluation Criteria for Uncertainty Handling:**

Score 0 - Deterministic Approach:

- The answer presents findings, predictions, or recommendations as certain facts

- No acknowledgment of uncertainty in climate projections, policy outcomes, or mechanisms

- Uses definitive language without qualification ("will happen", "certainly", "definitely")

Score 1 - Acknowledges Uncertainty:

- The answer mentions that there are uncertainties in:

* Climate mechanisms or pathways

* Model projections or scenarios

* Analysis results or data limitations

- However, uncertainty is mentioned in passing rather than being integrated into the reasoning

Score 2 - Uncertainty Integrated into Analysis:

- The answer not only acknowledges uncertainty but actively incorporates it into the analysis

- Provides robust strategies that account for uncertainty (e.g., scenario planning, robust decision-making, hedging strategies)

- Discusses confidence levels, ranges, or alternative futures

- Offers recommendations that are resilient across different uncertain outcomes

**Output Format:**

Provide ONLY a JSON object in the following format (no other text):

\{

"score": 0 or 1 or 2,

"reasoning": "Brief explanation of why this score was assigned (2-3 sentences)"

\}
\par\smallskip\hrule\smallskip\endgroup

\section{Evaluation of CFM as a foundation model for agents}

\subsection{Workflow}

\textbf{Evaluation framework.} We evaluate the agent capability of CFM using the ReAct (Reasoning and Acting) paradigm\numcite{69}, which interleaves reasoning and action during multi-step problem solving. This framework enables explicit modeling of intermediate decision processes and is therefore suitable for assessing agent-like behavior beyond final answer accuracy.

\textbf{Thought--Action--Observation cycles.} Model outputs are organized into iterative Thought--Action--Observation (TAO) cycles\numcite{70}. In each step, the model produces a \emph{Thought} to analyze the current problem state, selects an \emph{Action} corresponding to a specific operation, and receives an \emph{Observation} as the outcome. After multiple cycles, the model integrates the acquired information to produce a final answer.

A representative example is as follows:

\end{multicols}
\begingroup
\scriptsize
\renewcommand{\arraystretch}{0.86}
\setlength{\tabcolsep}{2pt}
\setlength{\LTleft}{0pt plus 1fill}
\setlength{\LTright}{0pt plus 1fill}
\begin{longtable}[]{@{}
  >{\raggedright\arraybackslash}p{(\columnwidth - 0\tabcolsep) * \real{1.00}}@{}}
\toprule
\textbf{Question:} What are the socioeconomic risks of transitioning away from fossil fuels in developing countries? \\
\midrule
\endhead
\textbf{Thought:} I need to understand the current role of fossil fuel industries in employment and economic structure.

\textbf{Action:} Query relevant data on fossil fuel-dependent communities

\textbf{Observation:} Fossil fuel industries account for approximately 5--10\% of employment in certain regions and contribute significantly to local economic activity

This structure allows us to trace intermediate reasoning, planning, and information integration throughout the problem-solving process. \\
\bottomrule
\end{longtable}
\endgroup
\begin{multicols}{2}

\textbf{Task data.} The evaluation is conducted on the same set of 220 climate tasks used in previous experiments, ensuring consistency with task-level evaluations. These tasks cover multidisciplinary climate problems and require varying levels of reasoning depth and knowledge integration.

\textbf{Tool action design.} To characterize tool-use behavior, we define six domain-relevant action types corresponding to common analytical patterns in climate science and policy analysis (Table I1). These actions collectively cover key operations in climate-related analysis, including information retrieval, mechanism understanding, quantitative reasoning, comparative analysis, impact assessment, and scenario projection.

\end{multicols}
\begin{center}
\textbf{Table I1. Tool action design}
\end{center}
\begingroup
\scriptsize
\renewcommand{\arraystretch}{0.86}
\setlength{\tabcolsep}{2pt}
\setlength{\LTleft}{0pt plus 1fill}
\setlength{\LTright}{0pt plus 1fill}
\begin{longtable}[]{@{}lll@{}}
\toprule
\textbf{Tool type} & \textbf{Function description} & \textbf{Example} \\
\midrule
\endhead
{[}Query{]} & Retrieve specific information from reports, databases, or literature & Retrieve carbon budget estimates from IPCC AR6 \\
{[}Analyze{]} & Conduct mechanism analysis or integrated assessment & Analyze renewable energy adoption trends in Southeast Asia \\
{[}Calculate{]} & Perform quantitative estimation or metric computation & Estimate emission reductions under a 2°C pathway \\
{[}Compare{]} & Compare policies, scenarios, or strategies & Compare carbon tax and emissions trading systems \\
{[}Evaluate{]} & Assess feasibility, risks, and impacts & Evaluate socioeconomic impacts of coal phase-out \\
{[}Forecast{]} & Project future trends or scenarios & Forecast renewable energy costs in 2035 \\
\bottomrule
\end{longtable}
\endgroup
\begin{multicols}{2}

\textbf{Simulated tool use.} We adopt a simulated tool-use setting, in which observations are generated by the model based on its internal knowledge rather than external APIs or databases. This design reduces variability from tool availability and data access, allowing the evaluation to focus on the model's intrinsic reasoning, planning, and tool-selection capabilities. While this setting does not fully replicate real-world deployment, it provides a controlled environment for assessing internal decision processes.

\textbf{Evaluation metrics.} We assess agent capability along two dimensions. \emph{Reasoning steps} measure the number of complete TAO cycles, reflecting the extent of multi-step reasoning and task decomposition. \emph{Action diversity} evaluates the range of distinct tool types used during reasoning, using a three-level scale (0--2) to capture the flexibility and functional richness of tool selection\numcite{71,72}. Together, these metrics quantify both the depth of reasoning and the breadth of action strategies in agent-like workflows.

\subsection{ReAct generation prompt}

The prompt for agent ReAct-based reasoning steps generation was as follows:

\begingroup\scriptsize\setlength{\parindent}{0pt}
\par\smallskip\hrule\smallskip
You are an expert climate policy and science advisor. You need to answer the following question using a ReAct (Reasoning + Acting) approach.

**ReAct Format Instructions:**

You should structure your response as a series of Thought-Action-Observation cycles, followed by a final Answer.

- **Thought**: Your reasoning about what information you need or what analysis to perform

- **Action**: The action you would take (e.g., "Query IPCC report on X", "Analyze policy case of Y", "Calculate emissions for Z", "Compare approaches A and B")

- **Observation**: What you would learn or observe from that action (simulate the result based on your knowledge)

Repeat Thought-Action-Observation cycles as needed (typically 3-5 cycles), then provide your final Answer.

**Example Format:**

Thought: I need to understand the current state of fossil fuel dependency in affected communities.

Action: Query economic data on fossil fuel employment and regional GDP contribution

Observation: Fossil fuel industries employ approximately 5-10\% of workforce in coal-dependent regions, contributing 15-30\% to local GDP...

Thought: Now I need to examine successful transition cases.

Action: Analyze just transition policies in Germany and Canada

Observation: Germany's coal phase-out includes €40 billion structural support fund...

{[}Continue with more cycles as needed{]}

Answer: {[}Your comprehensive final answer{]}

**Question:**

\{question\}

**Important:**

- Use at least 3 Thought-Action-Observation cycles

- Make your Actions diverse (query data, analyze cases, compare approaches, calculate impacts, etc.)

- Keep each cycle focused and coherent

- Your final Answer should synthesize insights from all cycles

Now, please answer the question using the ReAct format:
\par\smallskip\hrule\smallskip\endgroup

\subsection{ReAct evaluation prompt}

The prompt for agent evaluation was as follows:

\begingroup\scriptsize\setlength{\parindent}{0pt}
\par\smallskip\hrule\smallskip
You are an expert evaluator assessing the quality of ReAct-style agent reasoning.

**Question:**

\{question\}

**Model's ReAct Response:**

\{react\_response\}

**Evaluation Task:**

Evaluate this response across three dimensions:

1. **Reasoning Steps**: Count the number of complete Thought-Action-Observation cycles

- A complete cycle must have all three components (Thought, Action, Observation)

- Partial cycles don't count

2. **Action Diversity**: Assess the variety of actions taken

- Score 0: All actions are similar/repetitive (e.g., only "query" or only "analyze")

- Score 1: Some variety but limited (2-3 types of actions)

- Score 2: Good diversity (4+ different types of actions like query, analyze, compare, calculate, evaluate, etc.)

3. **Coherence Score**: Evaluate the logical flow and quality

- Score 0: Incoherent - cycles don't connect logically, observations don't follow from actions

- Score 1: Partially coherent - some logical flow but with gaps or inconsistencies

- Score 2: Highly coherent - clear logical progression, observations match actions, builds toward answer

**Output Format:**

Provide ONLY a JSON object (no other text):

\{\{

"reasoning\_steps": \textless number of complete cycles\textgreater,

"action\_diversity\_score": 0 or 1 or 2,

"coherence\_score": 0 or 1 or 2,

"action\_types": {[}"list of distinct action types identified"{]},

"reasoning": "Brief explanation of your evaluation (2-3 sentences)"

\}\}
\par\smallskip\hrule\smallskip\endgroup
\end{multicols}
\end{document}